\documentclass[journal,twocolumn]{IEEEtran}
\IEEEoverridecommandlockouts

\usepackage{cite}
\usepackage{algorithm}
\usepackage{algpseudocode}
\usepackage{graphicx}
\usepackage{textcomp}
\usepackage{xcolor}
\usepackage{balance}
\usepackage{paralist}
\usepackage[export]{adjustbox}
\usepackage{syntax}
\def\BibTeX{{\rm B\kern-.05em{\sc i\kern-.025em b}\kern-.08em
    T\kern-.1667em\lower.7ex\hbox{E}\kern-.125emX}}

\usepackage{subcaption}
\usepackage{subfiles}
\usepackage[colorlinks=true,linkcolor=blue,urlcolor=blue,citecolor=blue]{hyperref}

\usepackage{tikz}
\usepackage{circuitikz,tikz-timing}
\usetikzlibrary{circuits.logic.US}
\usetikzlibrary{shapes,calc,backgrounds,petri,positioning,automata,er}
\tikzstyle{arc} = [draw,->,thick,>=stealth]
\tikzstyle{dashedarc} = [draw,thick,dashed]
\tikzstyle{dottedarc} = [draw,thick,dotted]
\tikzstyle{thinarc} = [draw,->]
\tikzstyle{partarc} = [draw,-,thick]
\tikzstyle{tplace1} = [place,thick,scale=.7]
\tikzstyle{tplace} = [place,thick,scale=1]
\tikzstyle{tellipse} = [ellipse,draw,thick,minimum height=1cm]
\tikzstyle{ttransition} = [transition,thick]
\tikzstyle{redplace} = [tplace,fill=red!30]
\tikzstyle{redtransition} = [ttransition,fill=red!30]
\tikzstyle{redarc} = [arc,color=red]
\tikzstyle{blueplace} = [tplace,fill=blue!30]
\tikzstyle{bluearc} = [arc,color=blue]
\tikzstyle{greenplace} = [tplace,fill=green!30]
\tikzstyle{greentransition} = [ttransition,fill=green!30]
\tikzstyle{greenarc} = [arc,color=green]
\tikzstyle{smstate} = [state,thick]
\tikzstyle{bi} = [<->,draw,thick]

\tikzstyle{process} = [draw,thick,ellipse,minimum width=2cm,minimum height=1cm]
\tikzstyle{data} = [draw,thick,rectangle,minimum width=2cm,minimum height=1cm]
\tikzstyle{pipe} = [rectangle,align=center,draw,thick,color=red,minimum width=1.8cm]

\tikzstyle{datas} = [rectangle,draw,thick,text width=2cm,minimum height=1.3cm]
\tikzstyle{register} = [rectangle,draw,thick,minimum height=2cm,text width=1cm,align=center]
\tikzstyle{register1} = [rectangle,draw,thick,minimum width=4cm]
\tikzstyle{combicloud} = [cloud,draw,thick,minimum height=2.5cm,aspect=1,text width=1cm,align=center]
\tikzstyle{operator} = [rectangle,draw,thick,minimum width=4cm, minimum height=2cm]
\tikzstyle{box} = [rectangle,align=center,draw,thick,text width=1.5cm,minimum height=1cm, minimum width=1.5cm]
\tikzstyle{box2} = [rectangle,align=center,draw,thick,text width=2cm,minimum height=2cm]
\tikzstyle{box3} = [rectangle,align=center,draw,thick,text width=5cm,minimum width=5cm]
\tikzstyle{box4} = [rectangle,align=center,draw,thick,text width=3cm,minimum width=3cm]
\tikzstyle{box5} = [rectangle,align=center,draw,thick,text width=1.7cm,minimum width=1.7cm]
\tikzstyle{box6} = [rectangle,align=center,draw,thick,text width=1.0cm,minimum height=.8cm, minimum width=1.4cm]
\tikzstyle{decision} = [diamond,align=center,draw,thick,text width=0.7cm]
\tikzstyle{boldtext} = [ font=\bfseries ]
\tikzset{flipflop REG/.style={flipflop,
    flipflop def={t2=D, t5=Q, cd=1, td={} }
    }}
\tikzset{nflipflop REG/.style={flipflop,
    flipflop def={t2=D, t5=Q, cd=1, td={}, nd=1 }
    }}
\tikzstyle{reg} = [flipflop REG]
\tikzstyle{nreg} = [nflipflop REG,thick]

\tikzstyle{block} = [align=center,draw,thick,rectangle,minimum width=3.0cm,minimum height=2cm]
\tikzstyle{smallblock} = [align=center,draw,thick,rectangle]

\tikzstyle{vertex} = [circle,draw,thick,fill=black!10,font=\tiny\bfseries]

\tikzstyle{erent} = [entity,thick,align=center]
\tikzstyle{errel} = [relationship,thick,align=center]
\tikzstyle{erprop} = [attribute,thick,align=center]

\tikzstyle{chip} = [dipchip, num pins=6, hide numbers, external pins width=0.1, external pad fraction=64]

\tikzset{flipflop2 REG/.style={flipflop,
    flipflop def={t1=D, t6=Q, td=RST, c3=1, t3={\tiny{CLK}} }
    }}
\tikzstyle{reg2} = [flipflop2 REG]
\tikzset{nflipflop2 REG/.style={flipflop,
    flipflop def={t1=D, t6=Q, td=RST, c3=1, t3={\tiny{CLK}}, n3=1 }
    }}
\tikzstyle{nreg2} = [nflipflop2 REG]

\tikzstyle{latch1} = [
    flipflop,
    flipflop def={t2=D,td={\small{G}},t5=Q},
    ]
\tikzstyle{latch2} = [
    flipflop,
    flipflop def={t1=D,t3={\small{G}},td={\small{RST}},t6=Q},
    ]
\tikzstyle{gate} = [ draw, thick ]
\tikzstyle{xor} = [ xor gate US, gate ]
\tikzstyle{not} = [ not gate US, gate, scale=.7 ]
\tikzstyle{mullerc} = [ and gate US, gate, label=center:C ]
\tikzstyle{delem} = [ rectangle,draw,thick,minimum height=.7cm,minimum width=.8cm ]

\tikzset{timing/x/.style={draw=black}}

\tikzset{
  buswidth/.style={
    decoration={
      markings,
      mark=at position 0.5 with {
        \node {/};
        \node[below=1pt,sloped=false] {\tiny #1};
      }
    },
    postaction={decorate}
  }
}

\usepackage{siunitx}
\usepackage{listings}
\lstdefinelanguage{aa}{
    morekeywords={$barrier, $storage,$signal,$pipe,$uint,$depth,$module,$in,$out,$is,$constant,$branchblock,$dopipeline,$fullrate,$merge,$entry,$loopback,$phi,$on,$zero,$endmerge,$volatile,$slice,$mux,$while,$concat,$bitcast,$null,$call,$guard,$forkblock,$place,$parallelblock,$if,$then,$endif},
    otherkeywords={:,>=,<=,==,&&,+,-,!=,[],<,>,=,!,^},
    basicstyle=\ttfamily\footnotesize,
    breaklines=true,
    showstringspaces=false,
    showspaces=false,
    morecomment=[l]{//},
    commentstyle=\rmfamily\itshape,
}

\lstdefinelanguage{props}{
    morekeywords={patomic,log,quit},
    keywordstyle=\color{blue},
    basicstyle=\ttfamily\footnotesize,
    showstringspaces=false,
    showspaces=false,
    morecomment=[l]{\%},
    breaklines=true,
    }

\lstdefinelanguage{console}{
    basicstyle=\ttfamily\footnotesize,
    breaklines=true
}

\usepackage{amssymb} %
\usepackage{amsfonts} %
\usepackage{amsmath} %
\usepackage{amsthm} %
\usepackage{mathtools} %

\newtheorem{theorem}{Theorem}%
\newtheorem{lemma}[theorem]{Lemma}
\newtheorem{definition}{Definition}%

\newcommand{\always}{\Box\,}
\newcommand{\eventually}{\Diamond\,}
\newcommand{\until}{\,\mathcal{U}\,}

\newcommand{\occ}[1]{\ensuremath{#1_{occ}}}
\newcommand{\pcmap}[1]{\ensuremath{\mathcal{M}(#1)}}

\newcommand{\pred}[1]{\ensuremath{\prescript{\bullet}{}{#1}}}

\newcommand\dmin[2]{\ensuremath{d_{min}(#1,#2)}}
\newcommand\dmax[2]{\ensuremath{d_{max}(#1,#2)}}
\newcommand\dmini[1]{\ensuremath{d_{min}(#1)}}

\newcommand\sreq{\texttt{sreq}\,}
\newcommand\sack{\texttt{sack}\,}
\newcommand\ureq{\texttt{ureq}\,}
\newcommand\uack{\texttt{uack}\,}
\newcommand\din{\texttt{din}\,}
\newcommand\dinU{\texttt{dinU}\,}
\newcommand\Din{\ensuremath{D_{in}}\,}
\newcommand\DinU{\ensuremath{D_{in}^u}\,}
\newcommand\dout{\texttt{dout}\,}
\newcommand\doutU{\texttt{doutU}\,}
\newcommand\Dout{\ensuremath{D_{out}}\,}

\newcommand\fop{\ensuremath{f_{op}}\,}
\newcommand\cdep{\circlearrowleft}

\newcommand{\dpw}{\ensuremath{D_{pw}\,}}
\newcommand{\dsreq}{\ensuremath{D_{sreq}\,}}
\newcommand{\dsack}{\ensuremath{D_{sack}\,}}

\newcommand{\vc}[1]{\ensuremath{#1_{vc}\,}}
\newcommand{\ckt}[1]{\ensuremath{#1_{ckt}\,}}

\newcommand{\Fd}{\ensuremath{F_d\,}}
\newcommand{\Fc}{\ensuremath{F_c\,}}
\newcommand{\Fcd}{\ensuremath{F_{cd}\,}}
\newcommand{\Fdc}{\ensuremath{F_{dc}\,}}
\newcommand{\Fbr}{\ensuremath{F_{br}\,}}
\newcommand{\Tn}{\ensuremath{T_{n}\,}}
\newcommand{\Te}{\ensuremath{T_{e}\,}}
\newcommand{\tn}{\ensuremath{t_{n}\,}}
\newcommand{\te}{\ensuremath{t_{e}\,}}
\newcommand{\ls}{\ensuremath{l_{s}\,}}
\newcommand{\lt}{\ensuremath{l_{t}\,}}
\newcommand{\ps}{\ensuremath{p_{s}\,}}
\newcommand{\pt}{\ensuremath{p_{t}\,}}
\newcommand{\treq}{\ensuremath{t_{req}\,}}
\newcommand{\tack}{\ensuremath{t_{ack}\,}}
\newcommand{\fsynth}{\ensuremath{\mathcal{F}_{synth}\,}}
\newcommand{\pckt}{\ensuremath{P_{ckt}\,}}
\newcommand{\ickt}{\ensuremath{I_{ckt}\,}}

\newcommand\trisemax{\ensuremath{t_{rise.max}}\,}
\newcommand\tfallmax{\ensuremath{t_{fall.max}}\,}
\newcommand\tsetupmax[1]{\ensuremath{t_{setup.max}(#1)}\,}
\newcommand\tholdmax[1]{\ensuremath{t_{hold.max}(#1)}\,}
\newcommand\ttransmax{\ensuremath{t_{trans.max}}\,}

\newcommand{\xnor}{\mathbin{\overline{\oplus}}}
\newcommand{\xor}{\mathbin{\oplus}}
\newcommand{\procname}[1]{\mbox{\textsc{#1}}}

\title{An ASIC Design Flow for High-Level Synthesis of Asynchronous Circuits}

\author{
  Mayuresh Warunjikar,
  Madhav Desai
  \thanks{Mayuresh Warunjikar, Department of Electrical Engineering, Indian Institute of Technology Bombay, Mumbai. Email mayuresh@ee.iitb.ac.in}
  \thanks{Madhav Desai, Department of Electrical Engineering, Indian Institute of Technology Bombay, Mumbai. Email madhav@ee.iitb.ac.in}
}

\newcommand{\keywords}[1]{
\begin{IEEEkeywords}
#1
\end{IEEEkeywords}
}
\let\mysection\section

\let\mysubsection\subsection

\let\mysubsubsection\subsubsection

\newcommand{\intropageno}{}

\begin{document}
\maketitle

\begin{abstract}
Asynchronous circuits have some benefits over their synchronous counterparts, such as lower idle power consumption, potentially lower latency, absence of clock distribution issues, among others. However, industry-scale electronic design automation (EDA) support, particularly for high-level synthesis (HLS) of asynchronous systems from imperative programs, remains limited. We present an approach to synthesize asynchronous circuits from imperative programs using the AHIR HLS framework. The framework models the control path in the form of a restricted form of Petri nets that are 1-safe and live by construction. In this work, we start with this representation in AHIR and generate asynchronous circuits with a delay insensitive controller, a single-rail data path with bundled-data interaction between the controller and the data path that resembles the micropipelines architecture. We formalize the timing constraints required for correct functioning of the generated circuit and show that those constraints are one sided (feasible), necessary and sufficient. We have also developed a static timing analyzer to check that the circuit meets these constraints. Our flow integrates seamlessly with off-the-shelf ASIC toolchains and standard synchronous cell libraries. To demonstrate our approach, we show an end to end implementation of an AES encryption algorithm and use post layout circuit simulation to verify the correctness and estimate the performance of the resultant circuit.

\end{abstract}


\mysection{Introduction} \label{Sec:introduction}
\intropageno

\mysubsection{About asynchronous circuits}

Synchronous circuits have long been the dominant paradigm in VLSI design. In synchronous circuits, state transitions happen at discrete time steps, synchronized by a global clock signal. Asynchronous circuits, instead, are characterized by absence of a clock and state transitions happen using control signal handshakes between communicating components.
A synchronous circuit design assumes that the clock signal reaches all the sequential elements synchronously. With VLSI technology scaling, the gate delays reducing and wire delays becoming significant, meeting this assumption is more challenging due to clock skew issue\cite{clockskew}. Distributing the clock with sufficient signal strength also puts a significant drain on the power source \cite{clkpower,scalingclkpower}. In situations where the devices have to spend a larger part of their life in waiting for an external event, such as in edge devices, this significantly reduces the energy efficiency of the devices.

Asynchronous circuits, preempt several of these issues by not having a globally distributed clock and in turn the notion of clock skews. In idle states, asynchronous circuits, consume only the leakage power. In synchronous circuits, clock and power gating partly addresses this concern. However there are trade-offs involved between the idle power consumption and the wake-up latency \cite{power-gating} and the power supply still needs to be designed for the peak clock load. Low idle power consumption is natural in asynchronous circuits without complex clock and power gating scheme and wake-up latency trade off. With energy efficiency becoming a significant objective for sustainable computing \cite{energypropcompute,sustainability}, these aspects make asynchronous circuits an attractive option for VLSI designers.

Besides, asynchronous circuits have lower electromagnetic interference due to absence of concentrated switching at a single frequency that the synchronous circuits have \cite{lowemi}. A related and active area of research in the security domain is whether this makes the side-channel attacks harder \cite{self-timed-security,secure-crypto-eda,dpa-qdi,side-channel-di}.

Asynchronous circuits also permit a more modular development of hardware components. In synchronous circuits the timing constraints are global in nature. The clock frequency is limited by the slowest performing combinational logic component in the circuit. In asynchronous circuits, it is possible to focus the optimization effort on most frequently used components since a slower component does not impose a time constraint on other components.

This also implies that, in synchronous circuits the latency is a function of the worst case combinational delay, while in asynchronous case it is that of the average case delay, which potentially results in reduction in overall latency.

However, asynchronous circuits do have some challenges.
Unlike synchronous circuits, asynchronous circuits must operate correctly in continuous time, avoiding spurious transitions of signals, called glitches or hazards. Hence these circuits are also more sensitive to crosstalk. With continuous time state changes, formal verification methods face a greater state space explosion.
In absence of a central synchronizing signal (clock), handshake logic is required between any two communicating components, which adds to the area and latency.
Detection of completion of a computation needs to be explicit. Approaches like dual-rail design address this but add to significant energy and area costs.
Historically, the EDA tool support for asynchronous circuits is thin. No asynchronous aware placement and routing tools, no standards (e.g. to express timing constraints) have evolved for asynchronous systems. There are only a few academic HLS frameworks. Standard cell libraries have evolved only for synchronous designs. Testing methods, too, are focused on synchronous designs.

There are several application areas, where the energy efficiency and low idle power consumption are highly desirable where asynchronous circuits may find applications. There are several studies that show that the battery life in Implantable Medical Devices (IMD) is the primary concern given the surgical intervention it needs to replace a battery \cite{power-imd,cmos-imd,brain-imd,cochlear-imd}. Environmental Wireless Sensor Networks (EWSN) have similar challenges in replacement of batteries and stricter energy budgets, besides large idle times \cite{env-sensors}. There are several use cases of edge devices that have to work with energy constraints, with event driven bursts of activity followed by large idle times, where asynchronous circuits may be useful. For instance, event cameras, which asynchronously send pixel data when the brightness changes, have sparse asynchronous activity, low energy budget and low latency requirement \cite{event-cameras}.

\mysubsection{Our approach}

The goals of this exploration are:
\begin{inparaenum}
\item To produce a flow from HLS design entry to physical layout of an asynchronous integrated circuit, leveraging off-the-shelf flows and cell libraries.
\item To prove that the flow produces a circuit that faithfully implements the design entry.
\item Through experiments, assess the feasibility of the flow and the performance of generated circuits on various dimensions.
\item Assess the class of applications suitable for such methodology.
\end{inparaenum}

To this effect, the approach of this exploration, in brief, is:
\begin{inparaenum}
\item AHIR is an HLS framework \cite{sameerthesis} that accepts design entry in the form of an imperative language and converts it into an intermediate representation called virtual circuit. Further, it synthesizes synchronous circuit RTL from the virtual circuits. Our approach to synthesis of asynchronous circuits begins at the virtual circuit representation in AHIR.
\item The virtual circuit represents the control path using a restricted form of 1-safe and live Petri nets \cite{petri}, a data path and the control - data linkages. Our asynchronous synthesis tool produces RTL code from the virtual circuits. We describe our synthesis approach in Section \ref{Sec:LogicSynth}.
\item We design the data path and the control - data interaction using a latch based design. We formulate the timing constraints for correct functioning of this circuit. To check if an implementation meets these constraints we have developed a static timing analysis tool using OpenSTA \cite{opensta} tool as a library. We describe these aspects of the physical design in Section \ref{Sec:physical-design}.
\item We prove the correctness of the synthesis approach by way of trace equivalence between the virtual circuits and the synthesized circuits, through many steps, including reasoning about hazards in the synthesized circuit in Section \ref{Sec:verification}.
\item One of our goals is to use off-the-shelf tools to the extent possible to implement asynchronous circuits.  We use OpenLane \cite{openlane} toolchain and SkyWater cell library \cite{skywater130pdk} to take the synthesized design to a physical design and use our timing analysis tool to verify the timing constraints. We demonstrate the experimental results based on post-implementation timing simulations and various metrics gathered from a case study (128-bit AES encryption) in Section \ref{Sec:results}.
\end{inparaenum}

\mysubsection{Related work}

Asynchronous design methodologies are often characterized by the delay models they use. The delay models reflect the assumptions about the gate and wire delays that the methodology makes.

Huffman \cite{huffman} introduced one of the early delay models called \emph{bounded-delay model}. The circuits built using this model are also known as \emph{Huffman circuits}. These circuits operate in what is called the \emph{fundamental mode}, which permits only one of the inputs to change at a time. Upper bound on the time the circuit takes to stabilize after an input changes is expected to be known. The gap between input changes has to be greater than this bound. While such fundamental mode assumption is very restrictive for a practical circuit, it is useful for reasoning individual components in a circuit.

Nowick et al. \cite{nowick91} described an approach called \emph{Burst Mode} circuits that borrow some characteristics of synchronous circuits and some of Huffman circuits. In these circuits, multiple inputs are allowed to change in \emph{bursts}. A local clock is used to hold any outputs that feed back to the combinational logic in a stable state while they compute the response to a burst of inputs.

Ivan Sutherland, in his Turing Award lecture \cite{sutherland1989}, introduced \emph{Micropipelines}, in the form of an elegant design for asynchronous elastic pipelines. The approach uses a \emph{two-phase bundled-data} convention. In two phase signaling, also called \emph{transition signaling}, a control signal event is said to occur when its logic level changes from either 0 to 1 or 1 to 0. In bundled-data approach the data flow is moderated by control signals such that the delays in data transmission are less than the delays in the control signal that signals the writing of that data to a state holding element.

\emph{Delay Insensitive (DI) circuits} is the most robust model of asynchronous circuits, introduced by Molnar et al. \cite{molnar1985} and formally described by Udding et al. \cite{udding1986}. It makes no assumptions about the delays of wires and gates in a circuit. But, Martin \cite{martin1990} made some profound observations that proved the class of circuits that can be implemented as entirely DI circuits is very limited. He introduced a class of circuits called \emph{Quasi Delay Insensitive (QDI) circuits} by extending the DI circuits with a notion of \emph{isochronic forks}. Variations such as \emph{extended isochronic forks} by Berkel et al. \cite{berkel95}, extend the idea further. Manohar and Martin \cite{rajit95} showed that QDI circuits are Turing-complete. Manohar and Moses \cite{rajit2017} refined Martin's approach to cover a wider class of circuits.

\emph{Speed Independent circuits}, first described by Miller \cite{miller1979}, make an assumption that wire delays are negligible but the gate delays can be arbitrarily large. This approach is similar to QDI circuits, since the assumption about wire delays amounts to assuming that all forks are isochronic.

\emph{Signal Transition Graphs (STG)} \cite{chu86} is a methodology for specifying asynchronous circuits. STGs are essentially Petri nets in which rising and falling transitions on a wire are modeled as separate Petri net transitions. There are constraints on the Petri net, to ensure that the rising and falling transitions on a signal strictly alternate and they do so only after the successor transition of a given transition has taken place, to ensure the stability.

\emph{Globally Asynchronous, Locally Synchronous (GALS)}, a term first used by Chapiro \cite{chapiro84}, refers to a design pattern that combines some benefits of both the synchronous and asynchronous paradigms.

Beerel \cite{beerel2010} and Myers \cite{myers2001} have covered asynchronous circuits comprehensively in their books. Hauck \cite{scotthauck}, provides a good survey of various delay models and design methodologies in asynchronous circuits.
Balsa \cite{balsa} and Act \cite{act} are some of the mature frameworks that provide a design flow for asynchronous circuits.

Our approach is largely inspired by the Micropipelines architecture. We model the asynchronous circuits with a delay insensitive controller and a bundled-data interaction between the control and the data path. But we cater to fairly arbitrary control flows derived from imperative programs using a restricted form of Petri nets. Unlike Micropipelines, our approach does not require cells specifically designed for asynchronous circuits; it works with off-the-shelf ASIC flows and cell libraries designed for synchronous circuits.

The notion of delay-insensitivity in our synthesis of controller circuits is in line with Martin's insights on the limitations of delay-insensitivity. However, Martin's formulation that models a circuit as a set of production rules, expressly excludes the use of arbiters or branching elements. We permit the use of branch elements in our formulation. Although the branch element itself is not a DI circuit, we show that its use in the controller does preserve the DI property of the controller circuit. This allows us to build controllers that are DI, yet model fairly arbitrary control flows from imperative programs.

Our approach does not involve the use of QDI model or isochronic forks, which have stringent equality constraint on timing delays on forked paths that re-converge. We only have relative, one sided timing constraints.

Like STGs, we, too, use Petri nets to model the control paths. However, STG model relates very closely to the circuit realization, while our Petri nets are more abstract. For instance, transitions in an STG Petri net model a rising or falling edge of a logic level in a circuit, while in our Petri nets, a transition corresponds to a logical event. Due to their abstract nature they can be used for synthesizing asynchronous as well as synchronous systems. Further, our Petri nets isomorphically map to the asynchronous circuits we synthesize. STGs on the other hand require solving the Complete State Coding (CSC) problem, involving global reachability analysis \cite{chu86}. Our Petri nets, despite being restrictive, are easy to build from imperative programs, for which AHIR HLS provides a pathway.

Different mathematical formalisms are used to model and reason about asynchronous circuits by different researchers. For example, Martin's work uses production rules approach, STGs use transition systems.  We use \emph{prime event structures} \cite{winskeles}, a formalism based on modeling the causality and conflict relation between events and a mapping of 1-safe Petri nets to them, to reason about the behavior of our specification and the synthesized circuits. We find that that a \emph{true concurrency} formalism like this is useful to model asynchronous circuits where state transitions are continuous, unlike the discrete transitions in synchronous systems.

Systems like Act and Balsa have matured over years. There is a certain overlap of goals of these systems and those of ours. Our emphasis is on use of an imperative language for design entry and on leveraging and being interoperable with off-the-shelf ASIC flows and cell libraries, primarily designed for synchronous systems. Also, our framework offers a choice of both synchronous and asynchronous paradigms without requiring any change in the design entry. This is particularly useful for exploratory and research work.

Some asynchronous circuit design methodologies use \emph{dual-rail} data encoding as an alternative to the bundled-data approach. A dual rail system uses two wires to encode one bit of data to help detect completion of a computation. This helps in building a robust DI circuit where achieving timing closure is easy. However these benefits come for a significant area and energy costs. In present work, we choose the bundled-data method which saves on area and energy consumption. However tuning of the delays to meet the bundled-data constraints requires some manual steps which our static timing analysis tool assists in.

\mysection{Methodology} \label{Sec:methodology}

\mysubsection{AHIR flow from design entry to RTL}

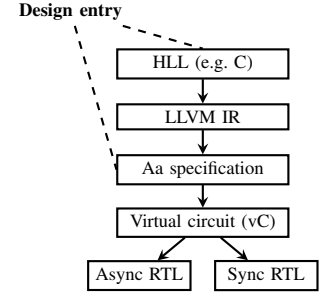
\begin{figure}[]
\centering
\begin{tikzpicture}[scale=.7,transform shape]
\node [boldtext] (de) at (-2.5,3) {Design entry};
\node [box4] (hll) at (0,2) {HLL (e.g. C)};
\node [box4] (llvm) at (0,1) {LLVM IR};
\node [box4] (aa) at (0,0) {Aa specification};
\node [box4] (vc) at (0,-1) {Virtual circuit (vC)};
\node [box5] (sync) at (1.2,-2) {Sync RTL};
\node [box5] (async) at (-1.2,-2) {Async RTL};
\path [dashedarc] (de) to (hll.north);
\path [dashedarc] (de) to (aa.west);
\path [arc] (hll) to (llvm);
\path [arc] (llvm) to (aa);
\path [arc] (aa) to (vc);
\path [arc] (vc) to (sync);
\path [arc] (vc) to (async);
\end{tikzpicture}

\caption{AHIR HLS flow from design entry to RTL}
\label{Fig:ahirflow}
\end{figure}
\begin{figure}[]
\centering

\subfloat[SSA pseudocode]{
\begin{minipage}{.14\textwidth}
\lstinputlisting[language=C]{example.c}
\end{minipage}
}
\hspace{.3cm}
\subfloat[Petri net]{
\begin{tikzpicture}[scale=.5,transform shape]
\node (branch) [tplace,label=right:branch] at(-2,2.5) {};
\node (fork) [ttransition,label=right:fork] at(0,1.5) {};
\node (s3req) [ttransition,label=right:s3req] at(-1,.6) {};
\node (s4req) [ttransition,label=right:s4req] at(1,.6) {};
\node (s3ack) [ttransition,label=right:s3ack] at(-1,-.6) {};
\node (s4ack) [ttransition,label=right:s4ack] at(1,-.6) {};
\node (join) [ttransition,label=right:join] at(0,-1.5) {};
\node (merge) [tplace,label=right:merge] at(-2,-2.5) {};
\path [arc] (-2,3.5) to (branch);
\path [arc] (branch) to (fork);
\path [arc] (fork) to (s3req);
\path [arc] (fork) to (s4req);
\path [dottedarc,->] (s3req) to (s3ack);
\path [dottedarc,->] (s4req) to (s4ack);
\path [arc] (s3ack) to (join);
\path [arc] (s4ack) to (join);
\path [arc] (join) to (merge);
\path [arc] (branch) to (merge);
\path [arc] (merge) to (-2,-3.5);
\end{tikzpicture}
}

\caption{An example of Petri net formation in AHIR from an SSA representation}
\label{Fig:aa2pn1}
\end{figure}
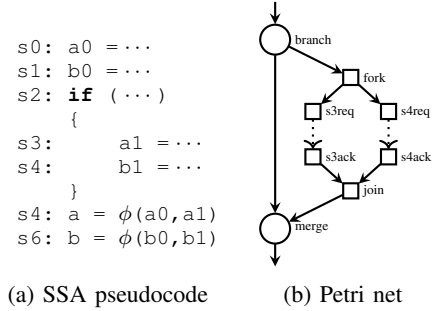

Fig. \ref{Fig:ahirflow} shows the design flow in AHIR high-level synthesis framework. This flow for synchronous circuits is described in detail in \cite{sameerthesis}. The steps in the flow are:
\begin{inparaenum}
\item AHIR has a design entry language named \emph{Aa}, which is an imperative language that predominantly uses single assignment variables \cite{cytron}. It has a notion of communication channels called \emph{pipes} and a notion of \emph{storage} which models the memory. The language has constructs to model pipelined flows in the form of loop statements. The language is fully described in \cite{aalrm}.
\item AHIR also accepts design entry in high-level languages, like C, that can be converted to an LLVM \cite{llvm} intermediate representation (IR). With reasonable restrictions such as absence of recursion and system calls.
\item The Aa specification is then converted into an intermediate representation called virtual circuit (vC), which also has a text form notation. The vC representation consists of the control path modeled as a Petri net, the data flow and the control-data interactions.
\item The vC representation is used to generate either synchronous or asynchronous RTL (in VHDL). The asynchronous synthesis process generates structural form VHDL code that instantiates components from our asynchronous RTL library.
\end{inparaenum}

Fig. \ref{Fig:aa2pn1} illustrates how the control flow aspect of an SSA form representation is modeled in a Petri net.\footnote{For brevity, we eliminate trivial (conduit) places and transitions.}, which shows four main aspects of a control flow - \emph{fork, join, branch} and \emph{merge}. The branches and merges modeled as places in Petri nets represent the conditional flows, while the forks and joins modeled as Petri net transitions represent the parallelism and synchronization. Each data path \emph{operation} follows a \emph{request-ack} protocol\footnote{To keep the illustration simple, we show only one request-ack pair per operator, though there are two such pairs per operator in AHIR virtual circuits.}, which are modeled as Petri net transitions. The dotted arcs indicate the dependencies between the request and ack transitions that are realized by the data path operators.

\mysubsection{Key components from asynchronous RTL library} \label{Sec:keycomponents}

\begin{figure}[]
\centering
\begin{circuitikz}[scale=.6,transform shape]
\node (tval) [anchor=east] at (-1,0.2) {truth-val};
\node (token) [anchor=east] at (-1,-2) {token};
\node (delay) [box] at (1,-2) {Delay};
\node (token2p) [box] at (3.5,-2) {T2P};
\node (tvand) [and gate US,draw,thick] at (6,0) {};
\node (tokand) [and gate US,draw,thick] at (6,-1.8) {};
\node (bubble) [ocirc] at ($ (tokand.west) + (0,.2) $) {};
\node (truepulse) [anchor=west] at (7,0) {true-pulse};
\node (falsepulse) [anchor=west] at (7,-1.8) {false-pulse};
\path [arc] (tval) to ($(tvand.west)+(0,.2)$);
\path [arc] (token) to (delay.west);
\path [arc] (delay) to (token2p.west);
\path [arc] (token2p.east) to ($(tokand.west)+(0,-.2)$);
\path [arc] (token2p.east) -- ($(token2p.east)+(.5,0)$) |- ($(tvand.west)+(0,-.2)$);
\path [arc] ($(tvand.west)+(-.5,.2)$) |- (bubble.west);
\path [arc] (tvand) to (truepulse);
\path [arc] (tokand) to (falsepulse);
\end{circuitikz}

\caption{Design of branch element}
\label{Fig:branchelem}
\end{figure}
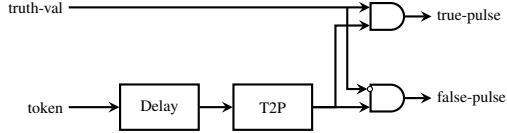

\begin{figure}[]
\centering
\begin{circuitikz}[scale=.6,transform shape]
\node [latch2] (latch) at (0,0) {};
\node [] (y) at ($(latch.pin 6)+(1,0)$) {y};
\node [] (vdd) at ($(latch.pin 1)+(-.5,0)$) {Vdd};
\node [and gate US,draw,thick] (and) at (-2,0) {};
\node [nor gate US,draw,thick,rotate=-90] (nor) at (-3,-1) {};
\node [or gate US,draw,thick] (or) at (-2,-2) {};
\coordinate (i0) at (and.input 1-|(-4,0);
\coordinate (i1) at (and.input 2-|(-4,0);
\coordinate (rst) at (or.input 2-|(-4,0);
\node [anchor=east] (i0l) at ($(i0)+(0,.1)$) {i0};
\node [anchor=east] (i1l) at ($(i1)+(0,-.1)$) {i1};
\node [anchor=east] (rstl) at (rst) {reset};
\path [arc] (latch.pin 6) to (y);
\path [arc] (and.east) -| (latch.pin 3);
\path [arc] (nor.output) |- (or.input 1);
\path [arc] (or.output) -| (latch.south);
\path [arc] (i0) to (and.input 1);
\path [arc] (i1) to (and.input 2);
\path [arc] (i0) -| (nor.input 1);
\path [arc] (i1) -| (nor.input 2);
\path [arc] (rst) to (or.input 2);
\end{circuitikz}

\caption{Design of Muller C element}
\label{Fig:mullercelem}
\end{figure}
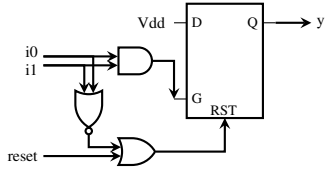

We describe two key components from the RTL library. Section \ref{Sec:LogicSynth} will provide mapping of Petri nets to the RTL components. Section \ref{Sec:advancedtopics} will describe some additional components.

Fig. \ref{Fig:branchelem} shows the design of a branch element that models a branch place, which uses an input from the data path to resolve the conflict. The T2P block converts a transition on the incoming Petri net arc into a pulse, which produces either a true-pulse or a false-pulse output, depending on the data path input. The transitions that succeed a branch place have a toggle flip flop, which flips its output value on receipt of the pulse, maintaining the transition signaling semantics. The delay element lets the truth value stabilize before interpreting it. In Section \ref{Sec:physical-design} we will be formalizing the constraints on the delay value and ways to meet those constraints.

Fig. \ref{Fig:mullercelem} shows a latch based design of a Muller C element that models a join transition. On system reset or when all inputs are at logic 0, the latch resets making the output logic 0. When all inputs are logic 1, the latch output becomes logic 1. When inputs are not in consensus, the output reflects the previously latched value. A hazard free operation requires that all the inputs change monotonically in one direction before any of them flips back, until the output changes. We will be proving this aspect in Section \ref{Sec:verification}.

\mysubsection{AHIR asynchronous RTL to implementation flow} \label{Sec:ahirasyncflow}

\begin{figure}[]
\centering
\begin{tikzpicture}[scale=.7,transform shape]
\node [box5] (lib) at (-1.2,2) {Async Lib};
\node [box5] (rtl) at (1.2,2) {Async RTL};
\node [box5] (delay) at (-2.2,1) {Delay spec};
\node [box5] (cell) at (2.2,1) {Cell spec};
\node [process] (synth) at (0,0) {Synthesis};
\node [box4] (synthnl) at (0,-1.25) {Synth Netlist};
\node [process] (openlane) at (0,-2.5) {OpenLane};
\node [box4] (implnl) at (0,-3.75) {Impl Netlist};
\node [process] (sta) at (0,-5.0) {STA};
\node [decision] (conscheck) at (0,-6.50) {Pass?};
\node [box4] (finalop) at (0,-7.75) {Final Outputs};
\node [boldtext, right=2pt of conscheck.south] {Y};
\node [boldtext, above=1pt of conscheck.west] {N};
\path [arc] (lib) to (synth);
\path [arc] (rtl) to (synth);
\path [arc] (delay) to (synth);
\path [arc] (cell) to (synth);
\path [arc] (synth) to (synthnl);
\path [arc] (synthnl) to (openlane);
\path [arc] (openlane) to (implnl);
\path [arc] (implnl) to (sta);
\path [arc] (sta) to (conscheck);
\path [arc] (conscheck) to (finalop);
\path [dashedarc,->] (conscheck) -|  (delay);

\end{tikzpicture}
\caption{ASIC flow for Asynchronous RTL generated}
\label{Fig:ahirasyncflow}
\end{figure}
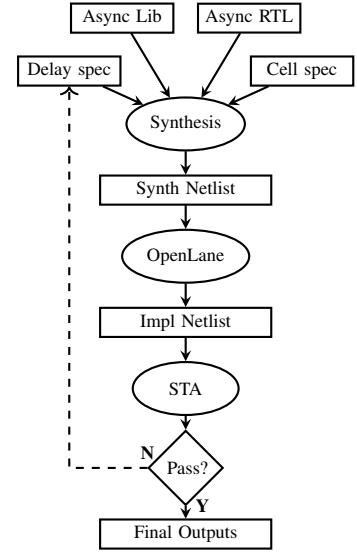

Fig. \ref{Fig:ahirasyncflow} shows the flow from RTL to the implementation. The generated RTL and the RTL library are fed to the synthesis process. The \emph{Cell spec} identifies the cell types that are used as delay elements, which are buffers that are specially designed to produce a combinational delay in an area efficient way. \emph{Delay spec} is a VHDL structure that is used to tune the delays inserted in the circuit to meet the timing constraints, illustrated in Fig. \ref{Fig:DelayModule}. The dashed arrow shows the iterative process of tuning the delay specification.

We use GHDL, an open source VHDL simulator, \cite{ghdl}, GHDL-Yosys plugin \cite{ghdl-yosys-plugin} and Yosys synthesis tool \cite{yosys} to convert these inputs into a synthesis netlist. We use the OpenLane \cite{openlane} framework and SkyWater open source PDK \cite{skywater130pdk} as the ASIC implementation flow. However, the flow can be easily adopted to work with other tools and PDKs. This flow produces an implementation netlist with placement, routing, parasitics information, SPICE model of the implementation etc.

\begin{figure}[]
\centering
\begin{lstlisting}[language=vhdl]
package vcdelays is
    constant nodedelays : t_delays_arr := (
 dp_1617 => (compute_delay => (3,0), others => (0,0)),
 -- ...
      others   => ( others => ( 1, 0 ) )
);
end package;
\end{lstlisting}

\caption{An example of delay specification}
\label{Fig:DelayModule}
\end{figure}

\begin{figure}[]
\centering
\begin{lstlisting}[language=console]
Note: unit=ps
Note: SLACK=ACTUAL-REQD
Note: -ve SLACK is FAILURE
Note: +ve SLACK is room for optimization
 TOPINST       DELAYTYP   ACTUAL     REQD    SLACK
 dp_1617  compute_delay  1631.80  2088.66  -456.86
...
\end{lstlisting}

\caption{Sample STA report}
\label{Fig:STAResult}
\end{figure}

The static timing analyzer (STA) in the flow, is developed by us for our asynchronous design methodology. We use the OpenSTA \cite{opensta} as a base tool to interpret the cell characterization, to read and interpret the parasitics information and to compute combinational delays in the circuit. Fig. \ref{Fig:STAResult} shows a sample output of the tool. In Section \ref{Sec:physical-design} we will describe the formulation of this delay tuning approach, its adequacy and feasibility. In Section \ref{Sec:futurework} we will list out some possible improvements to automate this approach and ways to arrive at better initial approximations for the delays to reduce the number of iterations.

\mysection{Logic synthesis from virtual circuits} \label{Sec:LogicSynth}

In this section we describe the virtual circuit representation with construction rules for the control path Petri nets, data path node behavior and how we synthesize an asynchronous circuit as an isomorphic mapping to structural HDL code.

\mysubsection{Construction rules of Petri nets used in AHIR}

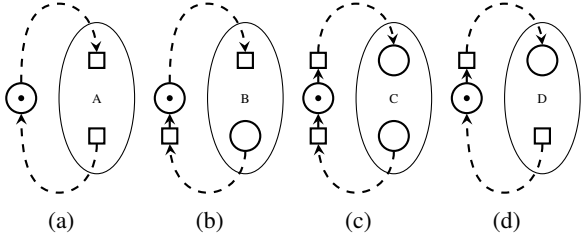
\begin{figure}[]
\centering

\subfloat[]{
\begin{tikzpicture}[scale=.5,transform shape]
\node (mp) [tplace,tokens=1] at (0,0) {};
\node (t1) [ttransition] at (2,1) {};
\node (t2) [ttransition] at (2,-1) {};
\node (l) [] at (2,0) {A};
\draw (2,0) ellipse(1cm and 2cm);
\path [arc,dashed]  (mp.north) to[out=90,in=180] (1,2.5) to[out=0,in=90] (t1.north);
\path [arc,dashed]  (t2.south) to[out=-90,in=0] (1,-2.5) to[out=180,in=-90] (mp.south);
\end{tikzpicture}
}
\subfloat[]{
\begin{tikzpicture}[scale=.5,transform shape]
\node (mp) [tplace,tokens=1] at (0,0) {};
\node (t1) [ttransition] at (2,1) {};
\node (t2) [tplace] at (2,-1) {};
\node (t3) [ttransition] at (0,-1) {};
\node (l) [] at (2,0) {B};
\draw (2,0) ellipse(1cm and 2cm);
\path [arc,dashed]  (mp.north) to[out=90,in=180] (1,2.5) to[out=0,in=90] (t1.north);
\path [arc,dashed]  (t2.south) to[out=-90,in=0] (1,-2.5) to[out=180,in=-90] (t3.south);
\path [arc,dashed]  (t3.north) to (mp.south);
\end{tikzpicture}
}
\subfloat[]{
\begin{tikzpicture}[scale=.5,transform shape]
\node (mp) [tplace,tokens=1] at (0,0) {};
\node (t1) [tplace] at (2,1) {};
\node (t2) [tplace] at (2,-1) {};
\node (t3) [ttransition] at (0,1) {};
\node (t4) [ttransition] at (0,-1) {};
\node (l) [] at (2,0) {C};
\draw (2,0) ellipse(1cm and 2cm);
\path [arc,dashed]  (t3.north) to[out=90,in=180] (1,2.5) to[out=0,in=90] (t1.north);
\path [arc,dashed]  (t2.south) to[out=-90,in=0] (1,-2.5) to[out=180,in=-90] (t4.south);
\path [arc,dashed] (mp.north) to (t3.south);
\path [arc,dashed] (t4.north) to (mp.south);
\end{tikzpicture}
}
\subfloat[]{
\begin{tikzpicture}[scale=.5,transform shape]
\node (mp) [tplace,tokens=1] at (0,0) {};
\node (t1) [tplace] at (2,1) {};
\node (t2) [ttransition] at (2,-1) {};
\node (t3) [ttransition] at (0,1) {};
\node (l) [] at (2,0) {D};
\draw (2,0) ellipse(1cm and 2cm);
\path [arc,dashed]  (t3.north) to[out=90,in=180] (1,2.5) to[out=0,in=90] (t1.north);
\path [arc,dashed]  (t2.south) to[out=-90,in=0] (1,-2.5) to[out=180,in=-90] (mp.south);
\path [arc,dashed] (mp.north) to (t3.south);
\end{tikzpicture}
}
\caption{Classification of token preserving regions (TPRs)}
\label{Fig:TPRs}

\end{figure}

\begin{figure}[]
\centering

\subfloat[]{
\begin{tikzpicture}[scale=.6,transform shape]
\node (t) [ttransition] at (0,0) {};
\node (p) [tplace] at (1.5,0) {};
\path [arc] (0,1) to (t.north);
\path [arc] (t.south) to (0,-1);
\path [arc] (1.5,1) to (p.north);
\path [arc] (p.south) to (1.5,-1);
\end{tikzpicture}
}
\subfloat[]{
\begin{tikzpicture}[scale=.6,transform shape]
\node (r0) [tellipse,minimum height=4cm,minimum width=1.5cm] at (0,0) {};
\node (r1) [tellipse] at (0,1) {};
\node (r2) [tellipse] at (0,-1) {};
\path [arc] (r1.south) to (r2.north);
\path [arc] (0,2.5) to (r1.north);
\path [arc] (r2.south) to (0,-2.5);
\end{tikzpicture}
}
\subfloat[]{
\begin{tikzpicture}[scale=.6,transform shape]
\node (t0) [ttransition] at (0,2.0) {};
\node (r1) [tellipse] at (-.5,1) {};
\node (r2) [tellipse] at (.5,1) {};
\node (t1) [ttransition] at (.5,-.5) {};
\node (t2) [ttransition] at (-.5,-1.5) {};
\node (r3) [tellipse] at (-.5,-3.0) {};
\node (r4) [tellipse] at (.5,-3.0) {};
\node (t3) [ttransition] at (0,-4.0) {};
\node (r0) [tellipse,minimum width=2.5cm,minimum height=7cm] at (0,-1) {};
\path [arc] (0,3.0) to (t0.north);
\path [arc] (t0) to (r1);
\path [arc] (t0) to (r2);
\path [arc] (r1) to (t2);
\path [arc] (r2) to (t1);
\path [arc] (t1) to (t2);
\path [arc] (t1) to (r4);
\path [arc] (t2) to (r3);
\path [arc] (r3) to (t3);
\path [arc] (r4) to (t3);
\path [arc] (t3) to (0,-5.0);
\end{tikzpicture}
}
\subfloat[]{
\begin{tikzpicture}[scale=.6,transform shape]
\node [tplace] (p0) at (0,3.0) {};
\node [tellipse] (r1) at (-.5,2.0) {};
\node [tellipse] (r2) at (.5,2.0) {};
\node [tplace] (p1) at (.5,.5) {};
\node [tplace] (p2) at (-.5,-.5) {};
\node [tellipse] (r3) at (-.5,-2.0) {};
\node [tellipse] (r4) at (.5,-2.0) {};
\node [tplace] (p3) at (0,-3.0) {};
\node [tellipse,minimum width=2.5cm,minimum height=7cm] (r0) at (0,0) {};
\path [arc] (0,4) to (p0);
\path [arc] (p0) to (r1);
\path [arc] (p0) to (r2);
\path [arc] (r2) to (p1);
\path [arc] (r1) to (p2);
\path [arc] (p1) to (p2);
\path [arc] (p2) to (r3);
\path [arc] (p1) to (r4);
\path [arc] (r3) to (p3);
\path [arc] (r4) to (p3);
\path [arc] (p3) to (0,-4);
\end{tikzpicture}
}
\caption[Type-2 Petri net construction rules]{Type-2 Petri net construction rules: (a) Primitive Regions (b) Series Region (c) Fork Region (d) Branch Region}
\label{Fig:Type2Rules}

\end{figure}
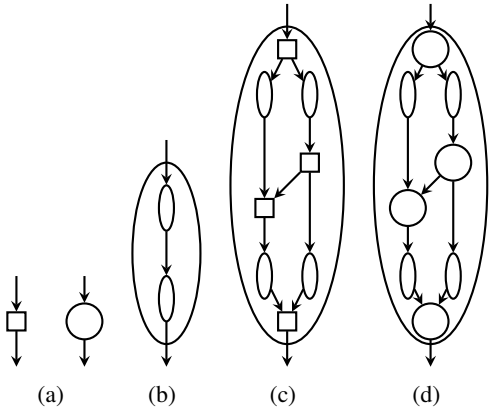

AHIR virtual circuits use \emph{Type-2 Petri nets} to represent the control path,  which are defined in \cite{sameerthesis}. We reproduce the same briefly for reference: A \emph{Type-1 Petri net} is a live and 1-safe Petri net that marks exactly one simple place in the initial marking.  A \emph{Token Preserving Region (TPR)} is a Petri net $P$ that can be augmented with one simple place $\hat{p}$ and a sufficient number of simple transitions and edges, to produce a live and 1-safe Petri net $P'$ such that $\hat{p}$ is the only marked place in the initial marking. A TPR has a unique \emph{entry} place or transition and a unique \emph{exit} place or transition. Depending on the type of entry or exit node, a TPR can be constructed in one of four ways called \emph{classes} of TPR as shown in Fig. \ref{Fig:TPRs}. The TPRs are unmarked 1-safe Petri nets. A Type-1 Petri net is constructed by placing a marked place around a TPR shown with dashed arcs in Fig. \ref{Fig:TPRs}. A \emph{Standard TPR (STPR)} is a TPR constructed using certain construction rules. These rules are shown in Fig. \ref{Fig:Type2Rules}. Finally, a \emph{Type-2 Petri net} is a Type-1 Petri net constructed using STPRs. Thus a Type-2 Petri net further restricts the Type-1 Petri nets by the construction rules of STPRs. \cite{sameerthesis} has shown that the Petri nets constructed using these rules are 1-safe and live.

\mysubsection{Behavior of a data path node in AHIR virtual circuit} \label{Sec:DPBehavior}

Some notation:
\begin{inparaenum}
\item We denote the functionality of a data path operator as $ \Dout = \fop(\Din) $, where \Din represents the input on the input data arc, \fop represents a function that the data path operator implements, \Dout represents the output on the output data arc.
\item A \emph{producer} is an entity that provides input data to a data path node. For the input ports of a circuit the \emph{environment} acts as a producer. In all other cases another data path node acts as a producer.
\item A \emph{consumer} is an entity that reads data from a data path node. For the output ports of a circuit the environment acts as a consumer. In all other cases another data path node acts as a consumer.
\item The control path interacts with a data path node using an interconnect protocol. To describe the protocol we use the symbolic notation described in Table \ref{Tab:SymProps}.
\end{inparaenum}

\begin{figure}[]
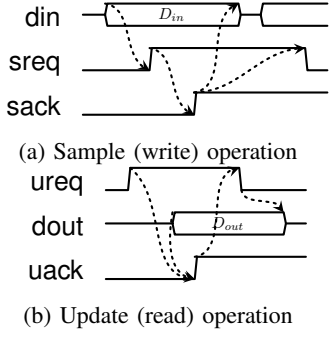

\centering

\subfloat[Sample (write) operation \label{Fig:protocol:write}]{
\begin{tikztimingtable}
din  & X 6D{\Din}   X    2D D \\
sreq & L 2L 2H 2H   H    2H L \\
sack & L 2L 2L 2H   H    2H H \\
\extracode
\begin{scope}[->, >=stealth, thick]
\draw [dottedarc,out=-10,in=160] (1,1) to (3,-2);
\draw [dottedarc,out=-10,in=160] (3.0,-1) to (5.0,-4);
\draw [dottedarc,out=10,in=200] (5.0,-3) to (7,1);
\draw [dottedarc,out=1,in=181] (5.0,-3) to (10,-1);
\end{scope}
\end{tikztimingtable}
} \hspace{1cm}
\subfloat[Update (read) operation \label{Fig:protocol:read}]{
\begin{tikztimingtable}
ureq & L 2H H 2H 2L   L \\
dout & X 2X 5D{\Dout} X \\
uack & L 2L L 2H 2H   H \\
\extracode
\begin{scope}[->, >=stealth, thick]
\draw [dottedarc,out=-40,in=179] (1.1,1) to (4,-4);
\draw [dottedarc,out=-100,in=175] (3.0,-1) to (4,-4);
\draw [dottedarc,out=10,in=185] (4,-3) to (6,1);
\draw [dottedarc,out=-80,in=100] (6.1,0) to (8,-1);
\end{scope}
\end{tikztimingtable}
}

\caption{Timing diagram of the data path interconnect protocol in AHIR virtual circuits}
\label{Fig:protocol}
\end{figure}

Each data path node in a virtual circuit has two interfaces:
\begin{inparaenum}
\item A \emph{write} or \emph{sample} interface, consists of a request event called \emph{sample request}, abbreviated \sreq and a \emph{sample ack} abbreviated \sack and a data input \Din. We symbolically represent availability of a valid input \Din on the data input arc by event \din. A data path node maintains a state in a register holding \Dout, by computing \fop(\Din). We denote an undefined value of the data input as \DinU and the event of its occurrence as \dinU. To ensure validity of the state, following rules are imposed:
    \begin{inparaenum}
    \item \label{writep1}  A producer must ensure that the data path node receives valid data before raising a request to write it i.e. $\din \cdep \sreq$.
    \item \label{writep2}  The data path node must issue an ack after completing the write operation. There must be exactly one ack issued after receiving a request. A producer must not raise a new request until an ack is received against an existing request i.e.  $\sreq \cdep \sack$
    \item \label{writep3}  A producer must ensure that the data input remains valid until it receives an \sack i.e. $\din \cdep \sack$
    \item Above property may be stated more explicitly as $[\sreq,\sack] \rightarrow \lnot\dinU$
    \end{inparaenum}
\item A \emph{read} or \emph{update} interface is similarly defined, which we omit for brevity. Fig. \ref{Fig:protocol} shows both the interfaces as timing diagrams.
\end{inparaenum}

In a pipelined design, the communication is initiated by the producer. Hence, on system reset, the initial \ureq event is implicitly triggered (described further in Section \ref{Sec:ploops}). Hence, any cyclic dependency of the form $\ureq \cdep E$, for some event $E$, flips to $E \cdep \ureq$. Also, the interval between \ureq and \uack where the data state must be valid becomes $[\uack,\ureq]$ since the first \ureq is raised after the first \uack in the pipelined mode. Thus:

\begin{definition} \label{Def:protocol}
\begin{inparaenum}
\item In the non-pipelined mode, the data path node's behavior for a given operator is defined by properties:
    \begin{inparaenum}
    \item \label{protocol:fnrel} Functional relation between the inputs and the outputs: $\Dout = \fop(\Din)$
    \item \label{protocol:safety} Safety properties specifying the event intervals in which the data must be in a valid state:
        \begin{inparaenum}
        \item $[\sreq,\sack] \rightarrow \lnot\dinU$
        \item $[\uack,\lnot\ureq] \rightarrow \lnot\doutU$
        \end{inparaenum}
    \item A set of cyclic dependencies that define the interconnect protocol: $\{ \din \cdep \sreq, \sreq \cdep \sack, \din \cdep \sack, \dout \cdep \uack, \ureq \cdep \dout, \ureq \cdep \uack \}$
    \end{inparaenum}
\item Similarly, in the pipelined mode:
    \begin{inparaenum}
    \item $\Dout = \fop(\Din)$
    \item \label{protocol:safety} Safety properties:
        \begin{inparaenum}
        \item $[\sreq,\sack] \rightarrow \lnot\dinU$
        \item $[\uack,\ureq] \rightarrow \lnot\doutU$
        \end{inparaenum}
    \item Cyclic dependencies: $\{ \din \cdep \sreq, \sreq \cdep \sack, \din \cdep \sack, \dout \cdep \uack, \dout \cdep \ureq, \uack \cdep \ureq \}$
    \end{inparaenum}
\end{inparaenum}
\end{definition}

\mysubsection{Synthesis as an isomorphic transformation of virtual circuit}

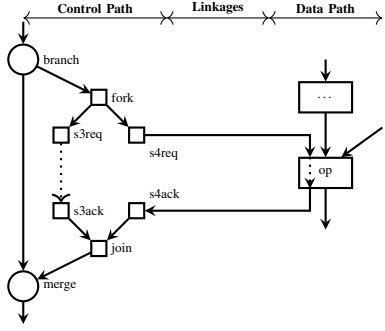
\begin{figure}[]
\centering

\begin{tikzpicture}[scale=.5,transform shape]
\node (branch) [tplace,label=right:branch] at(-2,3.0) {};
\node (fork) [ttransition,label=right:fork] at(0,2.0) {};
\node (s3req) [ttransition,label=right:s3req] at(-1,1) {};
\node (s4req) [ttransition,label=below right:s4req] at(1,1) {};
\node (s3ack) [ttransition,label=right:s3ack] at(-1,-1) {};
\node (s4ack) [ttransition,label=above right:s4ack] at(1,-1) {};
\node (join) [ttransition,label=right:join] at(0,-2.0) {};
\node (merge) [tplace,label=right:merge] at(-2,-3.0) {};
\node (s4) [box6] at (6.0,0) {op};
\node (din) [box6] at (6.0,2) {\ldots};
\coordinate (s4reqin) at ($(s4.north west)+(.3,0)$);
\coordinate (s4ackout) at ($(s4.south west)+(.3,0)$);
\coordinate (s4din2) at ($(s4.north east)+(-.3,0)$);
\path [arc] (-2,4.0) to (branch);
\path [arc] (branch) to (fork);
\path [arc] (fork) to (s3req);
\path [arc] (fork) to (s4req);
\path [dottedarc,->] (s3req) to (s3ack);
\path [arc] (s3ack) to (join);
\path [arc] (s4ack) to (join);
\path [arc] (join) to (merge);
\path [arc] (branch) to (merge);
\path [arc] (merge) to (-2,-4.0);
\path [arc] (din) to (s4);
\path [arc] (6,3) to (din);
\path [arc] (7.5,1.2) to (s4din2);
\path [arc] (s4) to (6,-1.5);
\path [arc] (s4req) -| (s4reqin);
\path [arc] (s4ackout) |- (s4ack);
\path [arc,dottedarc] (s4reqin) to (s4ackout);
\coordinate (cp0) at (-2,4.1);
\coordinate (cp1) at (1.8,4.1);
\coordinate (dp0) at (4.5,4.1);
\coordinate (dp1) at (7.5,4.1);
\draw [<->] (cp0) -- (cp1) node[midway,above,font=\bfseries] {Control Path};
\draw [<->] (cp1) -- (dp0) node[midway,above,font=\bfseries] {Linkages};
\draw [<->] (dp0) -- (dp1) node[midway,above,font=\bfseries] {Data Path};
\end{tikzpicture}

\caption{An example of control path, data path and the linkages in AHIR}
\label{Fig:aa2pn2}
\end{figure}

Fig. \ref{Fig:aa2pn2} shows a unified view of control and data path in AHIR virtual circuits. It expands one of the dotted arcs in Fig. \ref{Fig:aa2pn1} to show the control - data path linkages. The ingredients of this view are:
\begin{inparaenum}
\item A control path represented by a Type-2 Petri net.
\item A data path, a data flow graph where the nodes represent instances of operators and arcs represent the flow of data between them.
\item Control - data linkages with two request - ack pairs, for write (sample) and read (update) operations.
\end{inparaenum}

\begin{definition}\label{Def:vcdef}
AHIR virtual circuit is a graph $\vc{G} = (\vc{N},\vc{E},\vc{I})$ such that:
\begin{inparaenum}
\item $P$ is a set of unmarked places,
 $T$ is a set of transitions,
 $\Fc  \subseteq (P \times T) \cup (T \times P)$ is a flow relation between places and transitions in the control path
\item $D$ is a set of data path operator instances in the data path,
 $\Fd \subseteq (D \times D)$ is a data flow relation between the instances,
 $\Fcd \subseteq (T \times D)$: control - data interaction arcs that carry the \emph{request} signals
 $\Fdc \subseteq (D \times T)$: data - control interaction arcs that carry the \emph{ack} signals
 $\Fbr \subseteq (D \times P)$: data path input to place nodes, used for conflict resolution
\item $\Tn = \{place,transition,data\}$: a set of types that identifies the membership of the nodes to sets $P,T,D$ respectively.
\item $\Te = \{control,data\}$ is a set of edge type. Edges from sets \Fc, \Fcd and \Fdc are regarded as control edges and those from \Fd and \Fbr are regarded as data edges.
    \item $\vc{N} = P \cup T \cup D $,
    $\vc{E} = \Fc \cup \Fd \cup \Fcd \cup \Fdc \cup \Fbr$
    \item Each node $N \in \vc{N}$ is tuple $(l,\tn)$ where
        $l$ is unique label of $N$,
        $\tn\in \Tn$ is type of $N$
    \item Each edge $E \in \vc{E}$ that relates node $N_s$ and $N_t$ is tuple $(\te,\ls,\ps,\lt,\pt)$ such that:
        $\te\in \Te$ is type of $E$,
        $\ls = N_s.l$ and $\lt = N_t.l$ identify the source and target nodes by their labels respectively,
        $\ps,\pt \in \{0,1,2\ldots\}$ identify the ordinal position of the edge for a given edge type within the source and target node respectively.
    \item \vc{I} is a set of request-ack pairs $(\treq,\tack)$ associated with an operator instance. Virtual circuit semantics require that there be a one safe and live Petri net dependency between \treq and \tack, implemented by the operator. Each operator instance has two such request-ack pairs at ordinal positions 0 and 1 i.e.
        $\vc{I} = \{ (req(d,i),ack(d,i)) \mid d \in D, i \in \{0,1\}\}$ where
        $req(d,i) = \iota t\;(t\in T,e\in\Fcd,e.\lt=d.l,e.\pt=i,e.\ls=t.l)$,
        $ack(d,i) = \iota t\;(t\in T,e\in\Fdc,e.\ls=d.l,e.\ps=i,e.\lt=t.l)$
    \end{inparaenum}
\end{definition}

The synthesis process generates code in structural form, which contains instantiation of hardware components from a predefined library of component types and interconnections between such component instances. Thus, structural code can be viewed as a directed graph, whose nodes are component instances and edges are the interconnections between their ports. We define the synthesized code as:

\begin{definition}\label{Def:cktdef}
The synthesized code is graph $\ckt{G} = (\ckt{N},\ckt{E},\ckt{I})$, where the elements of the graph \ckt{N}, \ckt{E} and \ckt{I} follow the same annotation structure as \vc{G} in Definition \ref{Def:vcdef}. The role of each member of the annotation structure in the generated VHDL code is:
\begin{inparaenum}
\item Each node $N \in \ckt{N}$ is tuple $(l,\tn)$ where:
    $l$, the unique identifier of $N$, is used as the unique id of the generated component instance, 
    $\tn\in \Tn$ is passed as a compile time (generic) parameter with the instantiation of the component.
\item Each edge $E \in \ckt{E}$ is tuple $(\te,\ls,\ps,\lt,\pt)$ where:
    $\te\in \Te$, the arc type, identifies the component ports to connect,
    $\ls = N_s.l$ and $\lt = N_t.l$ identify the source and target component id respectively,
    $\ps,\pt \in \{0,1,2\ldots\}$ identify the ordinal position of the pin within a port to connect.
\item Implicit relation $I \in \ckt{I}$, has no explicit representation in the synthesized code. The operator library implements this relation as defined as specified in Definition \ref{Def:vcdef}.
\end{inparaenum}

\end{definition}

\begin{definition} \label{Def:fsynth}
Synthesis function is $\fsynth : \vc{G} \mapsto \ckt{G}$, such that:
$ \ckt{N} = \{ (N.l,N.\tn) \forall N \in \vc{N} \}$,
$ \ckt{E} = \{ (E.\te,E.\ls,E.\ps,E.\lt,E.\pt) \forall E \in \vc{E} \}$,
$ \ckt{I} = \{ (I.\treq.l,I.\treq.\tn), (I.\tack.l,I.\tack.\tn)) \forall I \in \vc{I} \} $
i.e. there is an identity mapping between the nodes, edges and event labels from \vc{G} to \ckt{G}.
\end{definition}

\mysubsection{Mapping of Petri net to asynchronous circuit} \label{Sec:Mapping}

\newcommand\construct[1]{\multicolumn{4}{l}{\textbf{#1}}\\\hline}
\newcommand\coltitle[1]{\multicolumn{1}{c}{\textbf{#1}}}
\begin{table*}[]
\centering
\caption[Mapping of Type-2 Petri nets to digital circuit]{Mapping of Type-2 Petri nets to digital circuit and their behavioral equivalence}
\begin{tabular}{%
@{}c%
c%
@{}c%
@{}c%
@{}c%
@{}c%
@{}c@{}%
}
\hline
\coltitle{Petri net}            &
\coltitle{Gate}                 &
\coltitle{Petri net behavior}   &
\coltitle{Input}                &
\coltitle{Map}                  &
\coltitle{Gate behavior}        &
\coltitle{Constraints}          
\\
\coltitle{                        } & %
\coltitle{                        } & %
\coltitle{$(P)$                   } & %
\coltitle{$(I)$                   } & %
\coltitle{$(M)$                   } & %
\coltitle{$(G)$ (for $i$ in 0,1)  } & %
\coltitle{$(E)$                   }   %
\\\hline
\construct{Primitive nodes}
\input{tab-row1.tex}
\\\hline

\construct{Fork}
\input{tab-row2.tex}
\\\hline

\construct{Join}
\input{tab-row3.tex}
\\\hline

\construct{Branch}
\input{tab-row4.tex}
\\\hline

\construct{Merge}
\input{tab-row5.tex}
\\\hline

\end{tabular}
\label{Tab:Mapping}

\end{table*}

Table \ref{Tab:Mapping} shows the gates we use to map the Petri net constructed using the construction rules described in Fig. \ref{Fig:Type2Rules}.

We use the term \emph{steady state} to mean either the initial state of a gate or its state after producing an output, till next change in any of its inputs. We use the term \emph{usage cycle} to mean the time period when a steady state ends (with change in state of any of its inputs) and when a new steady state starts (by change in state of the output). In the context of a given usage cycle, we call the initial steady state as \emph{precondition} and the next steady state as the \emph{postcondition}.

In Table \ref{Tab:Mapping}, we use symbols $t_0, t_1, t_2$ to represent the event of transfer of a token over a Petri net arc. We describe the behavior of Petri net, in a usage cycle, in terms of these symbols as a Boolean formula and represent it as $P$.
We use symbols $a, b, c, d$ to represent wires in the digital circuit representation of the Petri net. Each Petri net arc maps to a wire in the digital circuit. We suffix the symbols that represent the wires with $_0$ to represent the precondition and $_1$ to represent the postcondition of a usage cycle. We describe the behavior of each gate as a Boolean formula in terms of these suffixed variables and represent it as $G$.
We represent a Petri net's behavior in a usage cycle, at the input of a node by a Boolean formula $I$.
Our mapping requires that in steady state a certain constraint over the state of its inputs holds. We represent this constraint as a Boolean formula $E$.

\mysection{From synthesized circuit to physical design} \label{Sec:physical-design}

For physical design we leverage off-the-shelf tools, except for the timing aspect. We formulate the timing problem for our approach and provide a solution to the same. Before that, we describe the problem of hazards in an asynchronous circuits.

\mysubsection{Hazards in the controller circuit} \label{Sec:Hazards}

In asynchronous circuits, a logic level transition on a control path wire represents an event. Hence any spurious transition may lead to malfunctioning of the circuit. Such spurious transition is called a \emph{glitch} or a \emph{hazard}. The problem is well studied in the literature \cite{Eichelberger,nowick1993automatic,myers2001}. We use the terminology used by \cite{nowick1993automatic} to introduce hazards. In Section \ref{Sec:hazards-proof} we will prove their absence in our synthesized circuits and state assumptions for the same.

\mysubsection{Control - data event dependencies as timing constraints} \label{Sec:Timing}

\begin{figure}[]
\centering

\begin{tikzpicture}[scale=.6,transform shape]
\node (r1) [reg] at (-4,0) {$s$};
\node (dpcloud) [combicloud] at (0,0) {DP LOGIC};
\node (r2) [reg] at (4,0) {$t$};
\node (cpcloud) [combicloud] at (-1.5,-3) {CP LOGIC};
\node (delay) [box] at (1.5,-3) {DELAY};
\node (a) [vertex,anchor=east] at (-5,-3) {A};
\node (b) [vertex,anchor=south] at ($(r2.pin 2)+(.5,.2)$) {B};
\node (c) [vertex,anchor=south] at ($(r2.south)+(0,.2)$) {C};
\node (d) [vertex,anchor=south] at ($(delay.north)+(0,0)$) {D};
\path [arc] (r1.pin 5) to (dpcloud);
\path [arc] (dpcloud) to (r2.pin 2);
\path [arc] (a) to (cpcloud);
\path [arc] (cpcloud) to (delay);
\path [arc] (a) -| (r1.south);
\path [arc] (delay) -| (r2.south);
\end{tikzpicture}

\caption{Formulation of the timing problem}
\label{Fig:STAFormulation}

\end{figure}
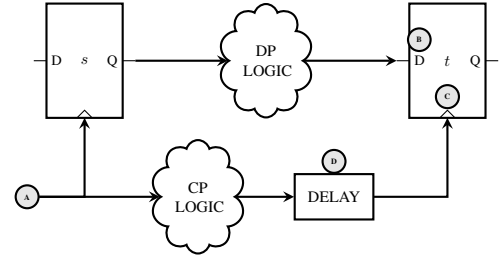

The event dependencies such as $\din \cdep \sreq$ (Definition \ref{Def:protocol}) require realizing the data path events like $\din$. In general, there are two ways to do this, \emph{dural rail} and \emph{bundled data}, which we discussed in Section \ref{Sec:introduction}. Present work explores the latter, which involves ensuring that the control signal to a state holding element reaches its control input pin sufficiently after a valid data input has reached its data input pin. Fig. \ref{Fig:STAFormulation} shows a representative circuit that depicts this timing problem.

\begin{inparaenum}
\item Let $S$ be the set of all state holding elements in the circuit.
\item Let data input pin of a state holding element be represented by $d$ and the control input pin by $c$.
\item Let $.$ denote the relation between a component instance and its pin, e.g. $s.d$ represents pin $d$ of component $s$.
\item Let \pred{} denote the driver of a pin i.e \pred{p} for an input pin $p$ denotes an output pin of some component that drives pin $p$.
\item Let $\rightarrow$ denote the data dependency between two state holding elements i.e. $s\rightarrow t$ denotes there is a combinational path from the data output pin of $s$ to the data input pin of $t$.
\item Let \dmin{a}{b} and \dmax{a}{b} denote the minimum and maximum combinational path delay between an output pin $a$ and an input pin $b$, respectively.
\item Let \dmini{D} represent the minimum combinational delay from the input to the output of a single-input single-output combinational component instance $D$.
\item Let $D_n$ represent a delay element responsible for meeting the timing constraints for component instance $n$.
\item Let \tsetupmax{e} represent the maximum setup time of the state holding element $e$.
\end{inparaenum}

Thus, for the circuit in Fig. \ref{Fig:STAFormulation}, $s$ and $t$ are state holding elements such that $s \rightarrow t$. The point marked $B$ represents pin $t.b$ and $C$ represents pin $t.c$. The point marked $A$ represents the driver pin for $s.c$ i.e. \pred{s.c}. Now, based on above discussion, the timing constraint that the circuit must meet is:
\begin{equation}\label{Eq:TimingConstraint}
\dmax{\pred{s.c}}{t.b} + \tsetupmax{t} < \dmin{\pred{s.c}}{t.c}
\end{equation}

To meet this constraint, we have to insert a \emph{delay element} $D_t$ (marked $D$ in Fig. \ref{Fig:STAFormulation}), which has to meet constraint:
\begin{equation}\label{Eq:DelayConstraint}
\dmini{D_t} > \dmax{\pred{s.c}}{t.b} + \tsetupmax{t} - \dmin{\pred{s.c}}{t.c}
\end{equation}

In general, the data input of a state holding element may be influenced by more than one source elements. However, each state holding element has only one delay element to meet its timing constraints. Thus. each delay element may have multiple such constraints. Notice that there is only a lower bound on \dmini{D_t}. Hence it is sufficient for $D_t$ to satisfy the constraint that demands the maximum delay. Hence, the set of all delay constraints is given by:
\begin{equation}\label{Eq:DelayConstraintSystem}
\begin{aligned}
\{ \dmini{D_t} > max(
    ( \dmax{\pred{s.c}}{t.b} + \tsetupmax{t} -\\
    \dmin{\pred{s.c}}{t.c}
    ) \forall s \in S | s \rightarrow t
    ) \forall t \in S \}
\end{aligned}
\end{equation}

Note that the system input and output ports can be trivially represented in the above formulation by modeling the environment as state holding elements.

\mysubsection{Data path interconnect design and timing constraints}\label{Sec:DPDesign}

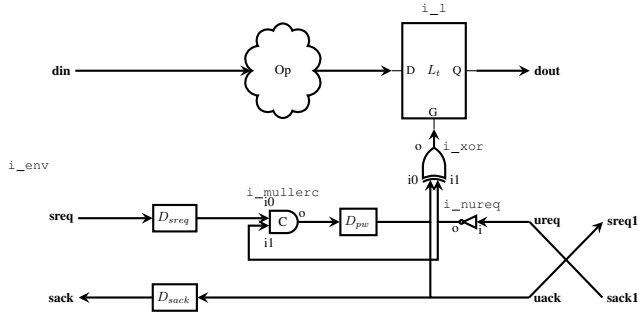
\begin{figure}[]
\centering
\begin{circuitikz}[scale=.5,transform shape]
\node (comb) [combicloud] at (0,0) {Op};
\node (lt) [latch1] at (4,0) {$L_t$};
\node (xort) [xor,rotate=90] at (4,-2.5) {};
\node (mullert) [mullerc] at (0,-4) {};
\node (dsreq) [delem] at ($(mullert.input 1)+(-2.5,0)$) {\dsreq};
\node (dpw) [delem] at (2.0,-4) {\dpw};
\node (notureq) [not,rotate=180] at (5,-4) {};
\node (ureq) [boldtext] at (7,-4) {ureq};
\node (uack) [boldtext] at (7,-6) {uack};
\node (sreq) [boldtext] at ($(dsreq)+(-3,0)$) {sreq};
\node (sack) [boldtext] at (uack-|sreq) {sack};
\node (dsack) [delem] at (dsreq|-sack) {\dsack};
\node (sreq1) [boldtext] at (9,-4) {sreq1};
\node (sack1) [boldtext] at (9,-6) {sack1};
\node (din) [boldtext] at (sreq|-comb) {din};
\node (dout) [boldtext] at (din-|ureq) {dout};
\path [arc] (din.east) to (comb.west);
\path [arc] (comb.east) to (lt.pin 2);
\path [arc] (xort.output) to (lt.down); 
\path [arc] (mullert.output) to (dpw.west); 
\path [arc] (dpw.east) -| (xort.input 1); 
\path [arc] (dsreq.east) to (mullert.input 1);
\path [arc] (notureq.output) -| (xort.input 2);
\path [arc] (ureq.west) to (notureq.input);
\path [arc] ( notureq.output -| xort.input 2) -- ++(0,-1) -- ++(-5,0) |- (mullert.input 2);
\path [arc] (sreq.east) to (dsreq.west);
\path [arc] (dsack.west) to (sack.east);
\path [partarc] (xort.input 1|-dpw.east) -- (xort.input 1|-uack) -- (uack);
\path [arc] (xort.input 1|-uack) -- (dsack);
\path [arc] (uack.west) to (sreq1.west);
\path [partarc] (sack1.west) to (ureq.west);
\path [arc] (lt.pin 5) to (dout);
\node [above=1mm] at(lt.north) {\texttt{i\_l}};
\node [above=2mm] at(mullert.north) {\texttt{i\_mullerc}};
\node [above=3mm] at(notureq.north) {\texttt{i\_nureq}};
\node [right=1mm] at(xort.east) {\texttt{i\_xor}};
\node [left=2mm] at(din|-xort) {\texttt{i\_env}};
\node [left=2mm] at(xort.input 1) {i0};
\node [right=2mm] at(xort.input 2) {i1};
\node [left=2mm] at(xort.output) {o};
\node [above=2mm] at(mullert.input 1) {i0};
\node [below=2mm] at(mullert.input 2) {i1};
\node at ($(mullert.output)+(0.1,0.2)$) {o};
\node at($(notureq.output)+(-.1,-.2)$) {o};
\node at($(notureq.input)+(.1,-.2)$) {i};
\end{circuitikz}

\caption{A pipeline stage}
\label{Fig:LatchStage}
\end{figure}

Fig. \ref{Fig:LatchStage} shows our data path interconnect design, as a pipeline stage. Pins $din, dout$ represent data input and outputs of the stage respectively. Pins $sreq, sack$ form a request - ack pair for the \emph{sample} (or write) operation, while pins $ureq, uack$ form it for the \emph{update} (or read) operation. $L_t$ is the latch that holds the state, which is computed by $Op$, a combinational logic that implements some operator.

\dsreq corresponds to element marked $D$ in Fig. \ref{Fig:STAFormulation}, meant to meet the constraints in Eq. \ref{Eq:DelayConstraintSystem}. \dsack delays the sample-ack signal to meet the hold time requirements of the latch. \dpw ensures that the Exor gate operates in fundamental mode, ensuring adequate gate pulse width. Thus, the hold time and internal functioning of the stage require the following timing constraints:
\begin{equation}\label{Eq:dsackc}
\begin{aligned}
\dmin{\texttt{i\_mullerc.o}}{\texttt{i\_env.sack}} > \\
\dmax{\texttt{i\_mullerc.o}}{\texttt{i\_l.G}} + \tholdmax{L_t}
\end{aligned}
\end{equation}
\begin{equation}\label{Eq:dpwc}
\begin{aligned}
\dmin{\texttt{i\_nureq.o}}{\texttt{i\_xor.i0}} > \\
\dmax{\texttt{i\_nureq.o}}{\texttt{i\_xor.i1}}
\end{aligned}
\end{equation}
Hence \dsack and \dpw have to meet the constraints:
\begin{equation}\label{Eq:dsack}
\begin{aligned}
\dmini{\dsack} > \dmax{\texttt{i\_mullerc.o}}{\texttt{i\_l.G}} +\\
\tholdmax{L_t} - \dmin{\texttt{i\_mullerc.o}}{\texttt{i\_env.sack}}
\end{aligned}
\end{equation}
\begin{equation}\label{Eq:dpw}
\begin{aligned}
\dmini{\dpw} > \dmax{\texttt{i\_nureq.o}}{\texttt{i\_xor.i1}} - \\
\dmin{\texttt{i\_nureq.o}}{\texttt{i\_xor.i0}}
\end{aligned}
\end{equation}

Note that the delays in Eq. \ref{Eq:dsack} and \ref{Eq:dpw} are not a function of the combinational logic delay of the data path operator. Empirically, on most typical implementation setups, the RHS of these two equations is much smaller than the gate delay of a single buffer cell. Hence, typically, introduction of a single delay buffer in place of \dsack and \dpw meets these two constraints.

\mysubsubsection*{Pipelined and non-pipelined usage of the interconnect}

As described with Definition \ref{Def:protocol}, in a pipelined mode the circuit raises the \ureq request implicitly on initialization. The circuit in Fig. \ref{Fig:LatchStage} uses inverter \texttt{i\_nureq} to raise the request implicitly. In a non-pipelined usage, the initial \ureq request has to be raised explicitly by a consumer. In such usage, the interconnect design replaces the inverter instance \texttt{i\_nureq} with a wire.

\mysubsection{Static timing analysis tool to impose the timing constraints} \label{Sec:statool}

Algorithm \ref{Fig:STAAlgo} describes our implementation that checks Eq. \ref{Eq:DelayConstraintSystem} holds.\footnote{We omit the details of support routines to conserve space. The names of support routines should give an idea about their functionality.}. The algorithm takes min and max corners, the parasitics information and implementation netlist as inputs . We build an abstract view of the netlist where state holding elements (flip flops and latches) and the delay elements we insert for meeting the timing constraints as the Timing Nodes.

\begin{algorithm}
\begin{algorithmic}

\Procedure{buildClksrcSet}{$REGS$}
    \State $clksrcSet \gets \emptyset$
    \For{ each \{ $r \in REGS \mid \text{ r is a data path reg}$ \} }
        \State $ r.clksrc \gets \Call{findDriver}{r.Clk} $
        \State insert $r.clksrc$ to $clksrcSet$
        \State $ r.clksrcClkMax \gets \Call{getMaxPathDelay}{r.clksrc,r.Clk} $
        \State $ r.clksrcQmax \gets  r.clksrcClkMax + r.tCQmax$
        \State $ r.clksrcClkMin \gets\Call{getMinPathDelay}{r.clksrc,r.Clk} $
        \State $ r.clksrcQmin \gets  r.clksrcClkMin + r.tCQmin$
    \EndFor
    \State \Return{$clksrcSet$}
\EndProcedure

\Procedure{buildClkDeps}{$clksrcSet$}
    \State /* $clkdepMap$ is a map of kind:
    \State $\{ clk \mapsto \{ clk \mapsto (DelayNode,C2CMinDelay) \} \}$ */
    \State $clkdepMap \gets \emptyset$
    \For{ each $csrc \in clksrcSet$ }
        \State $cdeps \gets \{ \text{nearest predecessor p} \mid p \in clksrcSet\}$
        \For{ each $cdep \in cdeps$ }
            \State $D \gets \text{DelayNode on path cdep to csrc}$
            \State \Call{assert}{D is unique}
            \State $Dmin \gets \Call{getMinPathDelay}{cdep,csrc} $
            \State /* C2CMinDelay is by excluding D's delay */
            \State $C2CMinDelay \gets Dmin - D.Dmin$
            \State $clkdepMap[csrc][cdep] \gets (D,C2CMinDelay)$
        \EndFor
    \EndFor
    \State \Return{$clkdepMap$}
\EndProcedure

\Procedure{STA}{}
\State \Call{initSTA}{}
\State $(REGS,DELAYS) \gets$ \Call{initNodes}{}
\State $clksrcSet \gets \Call{buildClksrcSet}{REGS}$
\State $clkdepMap \gets \Call{buildClkDeps}{clksrcSet}$
\For{ each \{ $t \in REGS \mid \text{ t is a data path reg}$ \} }
    \For{ each $s \in \Call{findDrivers}{t.D}$ }
        \State /* $(s,t)$ form a data dep pair */
        \State $AB \gets \Call{computeAB}{s,t}$
        \State $(d,AC) \gets \Call{computeAC}{s,t,clkdepMap}$
        \State $reqdDelay \gets AB - AC$
        \State \Call{requireDelay}{$d,reqdDelay$}
    \EndFor
\EndFor
\For{ each $d \in DELAYS$ }
    \State \Call{printDelayReport}{$d$}
\EndFor
\EndProcedure

\end{algorithmic}

\caption{STA algorithm}
\label{Fig:STAAlgo}
\end{algorithm}

We visit each cell instance once (\procname{initNodes}) to build the abstract timing node graph, then each timing node once (\procname{buildClksrcSet}) to identify the clock sources and each data dependency twice (in \procname{buildClkDeps} and \procname{STA}). Asymptotically, if $N,E$ represent the node and data dependency count of the netlist, then the complexity of this algorithm is, $O(N+E)$. The complexity of the complete process, however, is a function of the OpenSTA API complexity as well. We provide empirical data of the experiments we did for circuits with various gate, delay, register and data dependency counts, for reference.

\begin{table}[]
\centering

\caption{Run time of STA algorithm for various experiments}
\setlength{\tabcolsep}{4pt}
\begin{tabular}{@{}rrrrr@{}}
\hline
\textbf{GC}&\textbf{Regs}&\textbf{Delays}&\textbf{Deps}&\textbf{Runtime (s)}\\\hline
1969 & 469 & 52 & 394 & 2.85 \\\hline
3766 & 956 & 92 & 800 & 12.37 \\\hline
4057 & 253 & 58 & 896 & 12.59 \\\hline
4286 & 307 & 43 & 928 & 6.59 \\\hline
4411 & 361 & 40 & 960 & 7.26 \\\hline
5342 & 631 & 54 & 1120 & 10.22 \\\hline
6143 & 1412 & 246 & 728 & 4.73 \\\hline
15639 & 0 & 15 & 128 & 19.81 \\\hline
\end{tabular}

\label{Tab:STAMetrics}

\end{table}

\mysection{Verification} \label{Sec:verification}

Now we establish that the mapping from virtual circuits to the synthesized circuit is isomorphic, followed by the behavioral equivalence between the Type-2 Petri nets and the synthesized controller. Then we establish data path node behavioral equivalence under the formulated constraints and show that these constraints are feasible, necessary and sufficient. Lastly we prove the trace equivalence between the virtual circuit and the synthesized circuit.

\mysubsection{Isomorphism between virtual circuit and synthesized code} \label{Sec:isomorphic}

\begin{theorem} \label{Theorem:isomorphic}
Virtual circuit graph \vc{G} and the synthesized graph $\ckt{G} = \fsynth(\vc{G})$ (Definitions \ref{Def:vcdef}--\ref{Def:fsynth}) are isomorphic.
\end{theorem}
In graph \vc{G}, nodes \vc{N} are identified by a unique label. Nodes are annotated with type information. The implicit edges \vc{I} are identified by pairs of nodes. The edges \vc{E} are identified as a relation between two nodes and annotated with type and ordinal position information. \fsynth uses an identity function to map the node labels from \vc{G} to \ckt{G} and carries the additional structural information, such as node and arc types and arc ordinal positions to \ckt{G}.
Since an identity function is invertible, each element in \ckt{G} can be mapped back to its original element from \vc{G}, with the same type and ordinal position information. This shows that the function \fsynth is a bijection. Hence the graphs \vc{G} and \ckt{G} are isomorphic. \qed

\mysubsection{Equivalence between Type-2 Petri nets and their mapping}

\mysubsubsection*{Notion of equivalence between a Petri net and a digital circuit}

We represent flow of token over a Petri net arc by a transition in the logic level over the wire that represents the arc. For example, in Petri net, when we say transition $t_0$ occurs, in the digital circuit we say the wire $a$ representing its output arc, changes its logic level. Thus, the mapping $t_0 \mapsto a$ means, in a Boolean formula $t_0$ may be substituted with $a_0 \oplus a_1$. This substitution rule helps us formally assess the equivalence between a Petri net element and its corresponding digital circuit component.

\mysubsubsection*{On stability of the gate inputs}

\begin{lemma} \label{Lemma:InpDepOnOp}
In a circuit produced from Type-2 Petri nets, using mapping in Table \ref{Tab:Mapping}, for each gate, its input after an initial change causally depends on its output.
\end{lemma}

From the construction rules of Type-2 Petri nets, observe that:
\begin{inparaenum}
\item An STPR does not contain a marked place. A token has to enter the region via its unique entry point. Unless it exits the region via the unique exit point, it cannot re-enter the region.
\item Barring branch regions, other regions are acyclic. Hence, once a node receives a token, it can receive it again only after it leaves and re-enters the region.
\item A branch region permits cycles, but it disallows forks, which disallows replication of a token. And since there is no initial marking inside the region, there is at most one active token at any time in a branch region.
\item In the Petri net's mapping to a circuit, the flow of token is represented by a logic transition on a wire, which is responsible for change in input state of a gate.
\end{inparaenum}
Due to above restrictions, once an input of a gate undergoes a state change, next change in it is causally dependent on its output to change.
\qed

A causal dependence is necessary but not sufficient to guarantee that the inputs to a gate are stable until its output completes a transition. We will revisit this when we analyze hazards in the circuit Section \ref{Sec:Hazards}.

\mysubsubsection*{Outline of proof of equivalence between Type-2 Petri net and synthesized circuit}

At first we restrict our treatment to at-most two input gates and generalize it in Lemma \ref{Lemma:multi-input}. Recall the notation introduced in Section \ref{Sec:Mapping}.
\begin{inparaenum}
\item Basis the Type-2 Petri net construction rules, we prove the expressions $I$ and $E$ to be correct.
\item Using the notion of Petri net and digital circuit equivalence, we transform the Boolean formulas $P$ and $I$ in terms of the circuit variables, by using BDD composition, and call them \pckt and \ickt respectively.
\item Then we prove that the Petri net behavior \pckt and the gate behavior $G$ are equivalent if the constraint $E$ holds i.e. $ E \Rightarrow ( \pckt \Leftrightarrow G ) $
\end{inparaenum}

\begin{lemma}\label{Lemma:I}
Expression for input $I$ described in Table \ref{Tab:Mapping} is valid.
\end{lemma}

Given the Petri net is live by construction, except for merge node, all inputs of each node in the mapping table must receive a token in a usage cycle. Inputs of a merge place, on the other hand, have conflicting events on its inputs. Since the Petri net is 1-safe, in a given usage cycle, only one of the input transitions to a merge can occur. Hence Petri net behavior $I$ in the table is valid. \qed

We now compute \pckt and \ickt from $P$ and $I$ respectively, using the notion of equivalence and a BDD tool by composition as follows:

$\pckt = P[ t \mapsto (w_0 \xor w_1) \forall (t,w) \in M ] $

$\ickt = I[ t \mapsto (w_0 \xor w_1) \forall (t,w) \in M ] $

\begin{lemma}\label{Lemma:E}
Steady state of Petri net nodes satisfies their respective constraint $E$ in Table \ref{Tab:Mapping}.
\end{lemma}

We prove this argument by induction, by taking initial state as a base case and induction step as the postcondition obtained by applying input \ickt using characteristic function \pckt of the Petri net node, under the inductive assumption that the precondition meets the constraint $E$ and verifying that the postcondition, too, holds $E$.

Base case: In initial state of the circuit, the constraint $E$ holds for each gate.
Note that, primitive nodes, fork and merge are combinational circuits, whose constraint $E$ is same as gate behavior $G$. Hence the constraint holds for these node types. The branch and join nodes are sequential circuits. We assume that the circuit imposes the given constraint by way of initializing the state of the storage elements to 0 on system reset. Thus for all node types, in initial state the constraint holds.

Inductive step:
Given the Petri net node's characteristic function \pckt and input \ickt and under the inductive assumption that the constraint $E$ holds in the precondition, the postcondition $S_1$ can be computed as:
$S_1 = \exists(a_0,b_0,c_0,d_0)(\pckt \land \ickt \land E)$

To check if the constraint $E$ holds in postcondition $S_1$, we first express the constraint in terms of the postcondition variables i.e.
$ E_1 = E[a_0 \mapsto a_1\ldots d_0 \mapsto d_1] $

Now, if the postcondition meets the constraint, $ S_1 \Rightarrow E_1 $ must be true.
We check that this expression to be a tautology using BDD tool for all node types, which proves that the inductive step holds and hence constraint E holds in all steady states of a node. \qed

\begin{theorem}\label{Theorem:PNEquivalence}
Gate mapping described in Table \ref{Tab:Mapping} faithfully represents respective element in the Type-2 Petri net.
\end{theorem}

In Lemma \ref{Lemma:I} and \ref{Lemma:E} we established the correctness of $I$ and $E$ respectively. The Petri net behavior $P$ follows from the Petri net semantics, which is equivalently expressed as \pckt in terms of the circuit behavior and the gates are assumed to correctly implement the gate behavior $G$.  Using BDD tool we verify that $ E \Rightarrow ( \pckt \Leftrightarrow G ) $ is a tautology, which establishes that a node is faithfully represented by the gate it maps to in Table \ref{Tab:Mapping}, under constraint E which is shown to hold in a Type-2 Petri net.
\qed

\mysubsubsection*{Generalizing the proof for multi-input gates}

\begin{lemma} \label{Lemma:multi-input}
Theorem \ref{Theorem:PNEquivalence} holds in presence of multi-input joins and merges.
\end{lemma}

We can prove the identity $ E \Rightarrow ( \pckt \Leftrightarrow G ) $, by representing the expression $P,G,I$ and $E$ symbolically as input vectors of unknown size. We omit the proof for brevity.

\mysubsection{Hazards in controller circuit} \label{Sec:hazards-proof}

From mapping in Table \ref{Tab:Mapping}, the primitive nodes and forks map to wires and are trivially hazard free. The branch element receives an input from the data path for conflict resolution, involving a control - data interaction. We will be covering the treatment for such interaction in Section \ref{Sec:Timing}. We analyze the remaining two cases of joins and merges for potential hazards.

\textbf{Function hazards} :

\textbf{Case 1:} Exor function:
Per Lemma \ref{Lemma:I}, in one usage cycle, only one of the input transitions of a merge node can occur, which implies that only one of the inputs of the Exor gate can change i.e. the Exor gate operates in the \emph{fundamental mode}. If the initial state is $A$ and the changed state after the input transition is $B$, these states differ by exactly 1 bit. An Exor gate output changes when any one of its inputs changes i.e. $f(A) \neq f(B)$. Hence a static function hazard is absent. A dynamic function hazard requires a pair of intermediate states in the cube $[A,B]$. But since the states differ by only 1 bit, there are no intermediate states that can meet the criteria of the dynamic function hazard. Hence, the Exor function is free of function hazards in the context of the behavior of the merge in a Type-2 Petri net.

\textbf{Case 2:} Muller C function:
Per Lemma \ref{Lemma:E}, at the start of each usage cycle, the Muller C element inputs are in consensus. From the construction rules of a fork region in AHIR Type-2 Petri nets (Fig. \ref{Fig:Type2Rules}) notice that, a token cannot re-enter a region unless it has left the exit point of the region, which is a join transition for a fork region. Hence it is impossible for a join to receive a new token, unless the join transition triggers. Hence, the state change in the inputs of Muller C element is monotonic in a given usage cycle.

Consider a usage cycle that starts with all inputs of Muller C element in state logic 0, say state $A$. The usage cycle ends when all inputs flip their state to 1 - say state $B$. The cube $[A,B]$ contains several intermediate states where one or more, but not all of the input states are 1. Let's call such set of states $S_I$. The set of feasible transitions can then be described as:
\begin{inparaenum}
\item $ \{ A \rightarrow I \mid I \in S_I \}$
\item $ \{ I_1 \rightarrow I_2 \mid I_1,I_2 \in S_I, I_1 \neq I_2 \}$
\item $ \{ I \rightarrow B \mid I \in S_I \} $
\item $ \{ A \rightarrow B \} $
\end{inparaenum}

Note that an output state transition is possible only when the target state is $B$. Hence case 1 and 2 above need to be assessed for static function hazard and case 3 and 4 for dynamic function hazard. The only state that can cause an output transition is $B$, which is not a part of any transition cubes $[A,I]$ or $[I_1,I_2]$ in case 1 and 2. Hence static function hazard is not possible. Similarly, for case 3 and 4, no pair of states involving a state change is present in cubes $[I,B]$ or $[A,B]$ to cause a transition in the output. Hence a dynamic function hazard is not possible.

By symmetry, the above argument also applies when state $A$ has all inputs at logic 1 and $B$ has all inputs 0. Thus given the Type-2 Petri net construction rules and the initial state constraints, the Muller C element used as a join is free of function hazards.

\textbf{Essential hazards} :

In AHIR Type-2 Petri nets there are two ways to form a feedback path:
\begin{inparaenum}
\item For the top TPR (Fig. \ref{Fig:TPRs}) when a token leaves the TPR, it re-enters the TPR via a feedback loop. In the circuit, this feedback loop consists of an inverter gate.
\item In a branch region cyclic paths are permitted. A token can re-enter a cyclic region after it exits the region, via a merge node.
\end{inparaenum}

Per Lemma \ref{Lemma:InpDepOnOp}, every gate's every input causally depends on the gate's output after an initial change. However, a gate output takes certain finite duration to complete a transition. The loads it drives may possibly infer the logic level it is transitioning to, before it has stabilized. To avoid essential hazard, we have to ensure that this change does not propagate to the same gate's inputs via a feedback path.

Formally,
\begin{inparaenum}
\item Let \trisemax and \tfallmax represent maximum time for an output to rise from logic level 0 to 1 and fall from 1 to 0, respectively.
\item Then maximum time to complete a transition, $ \ttransmax = max(\trisemax,\tfallmax)$
\item Let $G$ be set of all gates and $I$ - a set of a gate's input pins, and $o$ a gate's output pin.
\item Let \dmin{a}{b} represent the minimum propagation delay from pin $a$ to pin $b$.
\item Then the timing constraint required to be met is:
$ \forall g \in G, \forall i \in g.I : \dmin{g.o}{i} > \ttransmax $
\end{inparaenum}

As there is only a lower bound on the delay, this constraint is feasible and can be easily met by adding a delay element, with delay value $> \ttransmax$ to the path for the two scenarios mentioned above.
Thus, the controller circuit is free of essential hazards, under the assumption that the feedback path around a TPR or via a merge node in a branch region has higher delay than the time required for a gate output to stabilize.
On a pragmatic note, the feedback path in AHIR Petri nets is never a self loop i.e. at least one gate is involved in the feedback path, with path delay far exceeding \ttransmax. Hence an explicit insertion of delay to preempt essential hazards is unnecessary.

\textbf{Logic hazards} :

Logic hazards are a characteristic of an implementation of the function. The implementations may vary depending on the synthesis tool chain and underlying logic cell implementations. We state assumptions about the underlying cell implementations for the Exor and Muller C elements.

\textbf{Case 1:} Exor implementation :
As shown before, the Exor gate in our controller operates in fundamental mode. Most of the implementation kits provide an Exor gate implementation that is hazard free in fundamental mode. It is typical for a cell library to include Exor gates up to 4 inputs. In situations where more inputs are involved, if the synthesizer produces a cascading of Exor gates, each gate in such cascaded arrangement operates in fundamental mode making the circuit free of logic hazards. However, it is imperative to examine the implementation to ascertain this.
With the assumption that the underlying Exor implementation is free of logic hazards in fundamental mode, our merge implementation is free of logic hazards.

\textbf{Case 2:} Muller C implementation :
Fig. \ref{Fig:mullercelem} shows our implementation of a Muller C element. As we analyzed in the function hazard, the possible input transitions are restricted, with all inputs changing monotonically from logic 0 to logic 1 in a usage cycle and back to logic 0 in an alternate cycle. As described in the case of Exor implementation, we make an assumption that the implementations of the logic gates are free of logic hazards. Given the set of possible transitions and this assumption, the gate and reset inputs to the latch are hazard free. Due to differing delays, a scenario of the gate acquiring logic 1 level while reset is yet to be logic 0, and the reverse of this can't be ruled out. However, in most latch implementations the reset gets precedence avoiding a race condition due to such scenario.
Thus assuming the gates used to be free of logic hazards and precedence for the reset signal in the latch implementation, our Muller C implementation is free of logic hazards.

\mysubsection{A Netlist to Promela converter tool}\label{Sec:nl2pml}

\begin{figure}[]
\centering

\begin{tikzpicture}[scale=.6,transform shape]
\node (nl2pml) [process] at (0,0) {nl2pml};
\node (nl) [anchor=east] at(-2,2.5) {netlist};
\node (cons) [anchor=east] at(-2,1.5) {constraints};
\node (lib) [anchor=east] at(-2,0.5) {gate library};
\node (env) [anchor=east] at(-2,-0.5) {environment model};
\node (props) [anchor=east] at(-2,-1.5) {symbolic properties};
\node (init) [anchor=east] at(-2,-2.5) {initial state};
\node (pml) [anchor=west] at(2,0) {promela code};
\path [arc] (nl.east) to (nl2pml);
\path [arc] (cons.east) to (nl2pml);
\path [arc] (lib.east) to (nl2pml);
\path [arc] (env.east) to (nl2pml);
\path [arc] (props.east) to (nl2pml);
\path [arc] (init.east) to (nl2pml);
\path [arc] (nl2pml) to (pml);
\end{tikzpicture}

\caption{Netllist2promela tool flow}
\label{Fig:nl2pml}
\end{figure}
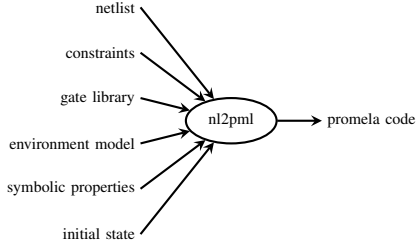

\begin{table}[]
\centering
\caption{Symbolic properties and their mapping to LTL}

\begin{tabular}{@{}p{.08\textwidth}@{}p{.10\textwidth}p{.25\textwidth}@{}}
\textbf{Symbolic property} & \textbf{Description} & \textbf{LTL formula} \\\hline

$\mathcal{L}(p)$ &
$p$ changes infinitely often &
$\lnot( ( \eventually \always p ) \lor ( \eventually \always \lnot p) )$ \\\hline

$\mathcal{L}_{all}$ &
$\mathcal{L}(p) \forall p \in P$ where $P$ is set of all pins &
$\lnot( ( \eventually \always p ) \lor ( \eventually \always \lnot p) ) \forall p \in P$
\\\hline

$a \circlearrowleft b$ &
$a$ and $b$ strictly alternate starting with $a$. Initial state: $a=b=0$ &
\vspace{-\baselineskip}
\[
\begin{aligned}
\always ( ( a \rightarrow ( a \until b))\\
\land (b \rightarrow ( b \until \lnot a))\\
\land (\lnot a \rightarrow ( \lnot a \until \lnot b ))\\
\land ( \lnot b \rightarrow ( \lnot b \until a)) )
\end{aligned}
\]
\\\hline

$[a,b]\rightarrow c$ &
Property $c$ holds in event interval $[a,b]$
&
$ \always ( ( a \rightarrow (c \until b) ) \land ( a \rightarrow ( \lnot a \rightarrow (c \until \lnot b)) ) )$
\\\hline

\end{tabular}

\label{Tab:SymProps}
\end{table}

To assist in building further proofs using SPIN model checker, we have developed a tool netlist2promela \cite{netlist2promela} to convert a circuit netlist description to Promela, the modeling language of the SPIN model checker. Fig. \ref{Fig:nl2pml} shows a context diagram of this tool.

The inputs to the tool are:
\begin{inparaenum}
\item \emph{netlist:} A custom json form to represent a circuit netlist.
\item \emph{constraints:} A model checker explores all interleavings of unordered events in a circuit. But the delay elements in a circuit are meant to preempt some of the interleavings. Our tool supports modeling the behavior of the delays by expressing a constraint on occurrence of an event on a given pin, until a given condition is met. The condition is expressed in terms of the pin states and a special variable \emph{val} that identifies the logic value that the event is about to set. Table \ref{Tab:NL2PMLConstraints}, shows an example of how a delay element is modeled with a pin and a constraint.
\item \emph{environment model:} The tool lets hand code additional processes in Promela, which is typically useful to model the environment, such as a producer and consumer component for the circuit under verification.
\item \emph{gate library:} Behavior of various gates that are used in the circuit and that of wires, is coded in Promela in this library.
\item \emph{symbolic properties:} The netlist2promela tool provides a symbolic notation for the properties one wishes to check. Table \ref{Tab:SymProps} describes the supported symbolic properties. The tool internally converts them into an LTL formulas. Further, our tool invokes another tool ltl2ba \cite{ltl2ba} to convert these properties into Promela code.
\item \emph{initial state:} describes the initial logic level of all the pins. The state is modeled using the \texttt{byte} data type in Promela. 3-valued logic (a symbolic value of 2 to mean an undefined value \DinU) can be used to define the initial state. The tool checks the consistency of these states with the gate behaviors and flags an error if the initial state is found to be inconsistent with the gate behaviors.
\end{inparaenum}

In the generated Promela code, each gate and wire is modeled as a concurrent process, which lets the model checker explore all possible interleavings, except those preempted by the constraints.

\mysubsection{Data path interconnect equivalence}\label{Sec:DPEquiv}

\mysubsubsection*{Verification setup}

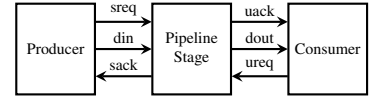
\begin{figure}[]
\centering
\begin{tikzpicture}[scale=.6,transform shape]
\node [box,minimum height=2cm] (stage) at (0,0) {Pipeline Stage};
\node [box,minimum height=2cm] (producer) at (-3,0) {Producer};
\node [box,minimum height=2cm] (consumer) at (3,0) {Consumer};
\path [arc] (producer.east) to node[midway,above] {din} (stage.west);
\path [arc] ($(producer.east)+(0,.6)$) to node[midway,above] {sreq} ($(stage.west)+(0,.6)$);
\path [arc] ($(stage.west)+(0,-.6)$) to node[midway,above] {sack} ($(producer.east)+(0,-.6)$);
\path [arc] (stage.east) to node[midway,above] {dout} (consumer.west);
\path [arc] ($(stage.east)+(0,.6)$) to node[midway,above] {uack} ($(consumer.west)+(0,.6)$);
\path [arc] ($(consumer.west)+(0,-.6)$) to node[midway,above] {ureq} ($(stage.east)+(0,-.6)$);
\end{tikzpicture}

\caption{Verification setup}
\label{Fig:Setup}
\end{figure}

Fig. \ref{Fig:Setup} describes our verification setup. We model the environment in the form of a \emph{producer} and a \emph{consumer} process. Both the producer and consumer run perpetually to help check that the properties of interest hold globally or infinitely often.
For the non-pipelined mode verification, the consumer process initiates the first \ureq request on initialization, while for the pipelined mode verification, the consumer waits for the \uack signal on initialization. In both modes, the consumer flips the request \ureq on receipt of a \uack, to start a fresh cycle.

The producer inserts the undefined value \DinU on the \din between receiving of an \sack and setting up of a new legitimate input on \din, to help check the properties related the validity of the output. To help check the liveness, and to be able to detect the data readiness as an event, the producer flips the \din value in each iteration.
We model the operator to be an identity function. Initial state is modeled by setting all the control inputs to logic 0 and data inputs to \DinU and setting the internal states consistently with these.

\begin{table}[]
\centering
\caption{Constraints that model the delay elements}

\begin{tabular}{@{}p{.06\textwidth}lp{.24\textwidth}@{}}
\hline
\textbf{Delay Element} & \textbf{Pin} & \textbf{Constraint}\\\hline

\dsreq &
$i\_l.G$ &
$val \lor (i\_env.din \xnor i\_env.dout)$
\\\hline

\dpw &
$i\_xor.i0$ &
$( i\_l.G \xnor ( i\_nureq.o \xor i\_xor.i0 ) )$
\\\hline

\dsack &
$i\_env.sack$ &
\vspace{-2\baselineskip}
\[
\begin{aligned}
( i\_l.G \xnor ( i\_nureq.o \xor i\_xor.i0 ) )\\
\land ( i\_mullerc.o \xnor i\_xor.i0 )
\end{aligned}
\]
\vspace{-\baselineskip}
\\\hline

\end{tabular}

\label{Tab:NL2PMLConstraints}
\end{table}

Table \ref{Tab:NL2PMLConstraints} describes how we model the delay constraints:
\begin{inparaenum}
\item \dsreq element's role is to enforce the timing constraint given by Eq. \ref{Eq:TimingConstraint}. The delay ensures that the \texttt{sreq} signal, which is responsible for closing the latch gate, reaches the latch only after the data has reached and reflected on the latch output. The constraint applies only when closing the gate, which is represented by the logical or with \emph{val}.
\item \dpw element's role is to ensure that both inputs of the Exor gate do not change simultaneously to ensure adequate gate pulse width. The input i0 of the Exor gate is constrained to reflect this delay.
\item \dsack delays the \texttt{sack} until the Muller C element's output first reaches the latch gate.
\end{inparaenum}

\mysubsubsection*{Verification of the data path interconnect}

With above setup, we verify the following results:

\begin{lemma}\label{Lemma:protocol}
If the constraints in Table \ref{Tab:NL2PMLConstraints} are satisfied and the environment (producer and consumer) follows the protocol Definition \ref{Def:protocol} for the respective usage modes, the pipeline stage design in Fig. \ref{Fig:LatchStage} is deadlock free and implements the behavior in Definition \ref{Def:protocol}.
\end{lemma}

Using above verification setup, we verify the properties $\mathcal{L}_{all}$ and all the properties in Definition \ref{Def:protocol}. \qed

\begin{lemma}\label{Lemma:necsuf}
The set of constraints in Table \ref{Tab:NL2PMLConstraints} is necessary and sufficient for the properties in Definition \ref{Def:protocol} to hold.
\end{lemma}

Using the model checking setup described, above properties are verified, which shows the sufficiency of the constraints. Dropping any one constraint invalidates one or more of the properties, which shows the necessity of these constraints. \qed

\begin{lemma}\label{Lemma:feasible}
The constraints in Table \ref{Tab:NL2PMLConstraints} are feasible.
\end{lemma}

The delays \dsack and \dpw are internal to the pipeline stage as they control relative delays between two internal paths of a pipeline stage and have only a lower bound  (Eq. \ref{Eq:dsack}, \ref{Eq:dpw}). The delay \dsreq is influenced by potentially multiple inputs that influence the combinational logic. As shown in Eq. \ref{Eq:DelayConstraintSystem} all constraints on this delay specify only a lower bound. Thus, all three constraints in Table \ref{Tab:NL2PMLConstraints} are feasible. \qed

\begin{lemma}\label{Lemma:implicit}
The data path interconnect in Fig. \ref{Fig:LatchStage} implements a 1-safe and live dependency between its requests and acks, as specified in Definition \ref{Def:vcdef}.
\end{lemma}
The properties $\sreq \cdep \sack$ and $\ureq \cdep \uack$ are verified using above verification setup, which implies a strict alternation between the respective request and ack pairs. The property $\mathcal{L}_{all}$ shows that every request is eventually acknowledged. Hence, per request exactly one ack is produced. This is equivalent of a 1-safe and live dependency between the request and its respective ack.
\qed

\begin{theorem}\label{Theorem:DataEquiv}
Circuit in Fig. \ref{Fig:LatchStage} faithfully represents a virtual circuit data path node for a given operator $Op$.
\end{theorem}

\begin{inparaenum}
\item We assume that the RTL library implements the operators correctly so that $ \Dout = \fop(\Din)$ holds.
\item Lemmas \ref{Lemma:protocol} - \ref{Lemma:feasible} show that the design in Fig. \ref{Fig:LatchStage} complies with the interconnect protocol (Definition \ref{Def:protocol}), under the timing constraints in Table \ref{Tab:NL2PMLConstraints}. Further, the constraints are shown to be feasible.
\item Lemma \ref{Lemma:implicit} shows that the design implements the implicit event dependencies that the virtual circuit specification (Definition \ref{Def:vcdef}).
\end{inparaenum}
Hence the design faithfully represents a virtual circuit data path node for a given operator.
\qed

\mysubsection{Trace equivalence between virtual and synthesized circuit} \label{Sec:Equiv}

We show that any event trace that the synthesized circuit produces belongs to a set of feasible traces in its specification, for a given initial state and input. We prove this is using Prime Event Structures \cite{winskeles} representation, which model the causality and conflict relations in a system.

\begin{definition}
Petri net is a tuple (B,E,F,M) where $B$ is a set of conditions (places), $E$ is a set of events (transitions), $F \subseteq (B \times E) \cup (E \times B)$ is flow or causal dependency relation (arcs) and $M \subseteq B$ is a marked subset of the conditions (places).
\end{definition}

A notion of \emph{unfolding} or \emph{occurrence net} of Petri nets was introduced by Ken McMillan \cite{mcmillanunfoldings}. Unfoldings are Petri nets that are acyclic and free of backward conflict (merge places), that represent causality and conflict relations in the original Petri net they are derived from.

\begin{definition}\label{Def:OccNet}
Given a 1-safe Petri net $P = (B_0,E_0,F_0,M_0)$ its occurrence net \occ{P} $= (B_1,E_1,F_1,M_1)$ is a 1-safe Petri net such that $F_1$ is acyclic and free of merges i.e. $ \forall b\in B_1 |\pred{b}| \le 1$. There is a mapping $\lambda$ from the places and transitions of \occ{P} to $P$. The flow relation $F_1$ is consistent with this mapping i.e. $\forall (a,b) \in F_1. (\lambda(a),\lambda(b)) \in F_0$. Initial marking $M_1$ is not preceded by any events in the relation $F_1$ i.e. $\forall b\in M_1. \pred{b} = \emptyset$
\end{definition}

McMillan \cite{mcmillanunfoldings} introduced a notion of configuration. A configuration represents a feasible \emph{partial run} of an occurrence net.

\begin{definition}\label{Def:configuration}
A configuration $C$ is a set of events of an occurrence net \occ{P} $= (B,E,F,M)$ such that $C \subseteq E$, $C$ is causally closed i.e. $e \in C \Rightarrow e' \in C \; \forall e' \le e$ where $e' \le e$ represents a causal relation (path) from $e'$ to $e$ and $C$ is conflict free i.e. $e_1,e_2 \in C \Rightarrow \lnot(e_1 \# e_2)$. Relation $\#$ represents a conflict relation over events. When two events share a predecessor, they are in (direct) conflict i.e. $\pred{e_1} \cap \pred{e_2} \neq \emptyset \Rightarrow e_1 \# e_2$. The events causally dependent on conflicting events are in (inherited) conflict i.e. $ e_1' \# e_2', e_1' \leq e_1, e_2' \leq e_2 \Rightarrow e_1 \# e_2$
\end{definition}

Let \vc{PN} represent a Petri net view of the graph \vc{G} (Definition \ref{Def:vcdef}) and \ckt{PN} represent the control path view of the synthesized graph \ckt{G} (Definition \ref{Def:cktdef}), both with explicit representation for the request-ack dependencies via the data path operators (Lemma \ref{Lemma:implicit}).  Let $\mathcal{M} : \vc{PN} \rightarrow \ckt{PN}$ represent the mapping between the representations. As we showed in Theorem \ref{Theorem:isomorphic} $\mathcal{M}$ is a bijection.

We prove the trace equivalence in two steps. First, we prove that the feasible configurations producible by either representation are in bijection. Second, we show that any event trace in the synthesized circuit is a subset of feasible traces in the virtual circuit, by linear expansion of configurations into traces.

\begin{lemma}\label{Lemma:CausalityPreserve}
Mapping $\mathcal{M}$ is causality preserving i.e. $n_1,n_2 \in \vc{PN}, n_1 \leq n_2 \Rightarrow n_1' \leq n_2' \text{ where } n_1' = \pcmap{n_1}, n_2' = \pcmap{n_2}$.
\end{lemma}

In a Petri net, except for a merge place, the arcs form a causal relation for a node. Per mapping in Table \ref{Tab:Mapping}, primitive nodes, forks and branches map to gates with a single input and their behavior $G$ shows that their output is causally dependent on the input. The joins map to Muller C element and given the constraint $E$, all incoming wires must undergo a transition for the output transition to trigger. Thus, except for the gate that the merge place maps to all wires represent a causal relation. A configuration is based on an unfolded occurrence net of the respective representation in which the merge node draws its inputs from one of its predecessors on which it is causally dependent. Thus the mapping $\mathcal{M}$ is causality preserving.  \qed

\begin{lemma} \label{Lemma:ConflictPreserve}
Mapping $\mathcal{M}$ is conflict preserving i.e. $n_1,n_2 \in \vc{PN}, n_1 \# n_2 \Rightarrow n_1' \# n_2' \text{ where } n_1' = \pcmap{n_1}, n_2' = \pcmap{n_2}$.
\end{lemma}

We introduced the terms \emph{direct conflict} and \emph{inherited conflict} in Definition \ref{Def:configuration}. As per the Petri net semantics, when a place node has more than one successors, they are in a direct conflict. Such branch place maps to a branch element. The behavior of the branch element ($G$ in Table \ref{Tab:Mapping}) shows that the successors of branch element are in conflict as only one of its output wires undergo a logic level transition, per configuration. Thus, the direct conflict relation is preserved in the mapping. Since the inherited conflict is a function of the causal relation, which is also preserved (Lemma \ref{Lemma:CausalityPreserve}), the mapping preserves the conflict relation.
\qed

\begin{lemma} \label{Lemma:ConfigBijection}
For any configuration \vc{C} produced in \vc{PN}, there exists a configuration \ckt{C} in \ckt{PN} such that \vc{C} and \ckt{C} are in bijection.
\end{lemma}

We prove this argument by inductively constructing \vc{C} using Definition \ref{Def:configuration} and showing that for every construction step of \vc{C}, there exists an equivalent step in \ckt{C}. Note that the configurations are sets, hence the sequence of triggering of events is not relevant.

\emph{Base case:}
Consider an empty configuration $\vc{Cpre} = \emptyset$ of \vc{PN} and $\ckt{Cpre} = \emptyset$ of \ckt{PN}. There exists a unique entry transition in a Type-2 Petri net. Liveness implies that the entry transition can be triggered. Equivalent entry transition can be triggered in \ckt{PN}. The new configurations after addition of this event are $ \vc{Cpost} = \{ \vc{en} \} $ and $ \ckt{Cpost} = \{ \ckt{en} \}$. By symmetry of the argument, $\vc{Cpost}$ and $\ckt{Cpost}$ are in bijection.

\emph{Inductive step:}
Given configurations \vc{Cpre} and \ckt{Cpre} which are in bijection (by inductive assumption), consider an event \vc{e} that can be added to \vc{Cpre}. Hence \vc{e} must meet two criteria in Definition \ref{Def:configuration}:
\begin{inparaenum}
\item Causally closed: i.e. $e' \leq \vc{e} \Rightarrow e' \in \vc{Cpre}$
\item Conflict free: i.e. $ e' \in \vc{Cpre} \Rightarrow \lnot(e' \# e)$.
\end{inparaenum}
But since, the causality and conflict relations are preserved through bijective mapping $\mathcal{M}$ (Lemma \ref{Lemma:CausalityPreserve} and \ref{Lemma:ConflictPreserve}) an equivalent event \ckt{e} must exist, that can be added to \ckt{Cpre}.
After these additions, $\vc{Cpost} = \vc{Cpre} \cup \{\vc{e}\}$ and $\ckt{Cpost} = \ckt{Cpre} \cup \{\ckt{e}\}$. \vc{Cpre} and \ckt{Cpre} are in bijection and there exists a mapping between \vc{e} and \ckt{e}. By symmetry of the construction process, the inverse mapping exists. Hence \vc{Cpost} and \ckt{Cpost} are in bijection.
From the construction rules \vc{PN} has a unique exit node and from the liveness of the Type-2 Petri nets, the exit node is reachable. Hence the configuration construction process terminates and a configuration is finite. By principle of structural induction, configurations of \vc{PN} and \ckt{PN} are in bijection.
\qed

Given a configuration, which is a causally closed and conflict free set of events, a set of feasible traces can be constructed by linear extension (also called linearization or topological sort). We define it as a linearization operator as follows:

\begin{definition}
For any configuration $C$, built using causal relation $\leq$, the set of feasible traces is given by linear expansion denoted as $Lin(C,\leq)$, such that $\forall T \in Lin(C,\leq)$, $T$ is a total order $\leq^*$ such that $\forall x,y \in T, x \leq y \Rightarrow x \leq^* y$.
\end{definition}

\begin{theorem}
An event trace produced by our synthesized circuit belongs to the set of feasible traces of the AHIR virtual circuit.
\end{theorem}

Let \ckt{C} be any configuration producible by the controller circuit \ckt{PN}. The set of feasible traces for this circuit is then $Lin(\ckt{C})$.
For any event trace $\ckt{T}$ produced by the circuit $\ckt{T} \in Lin(\ckt{C})$.
As per Lemma \ref{Lemma:ConfigBijection}, for the given configuration \ckt{C}, there must exist an equivalent configuration \vc{C} for \vc{PN}, with both configurations being in bijection. The set of feasible traces for this configuration is $Lin(\vc{C})$.
Since the causal relation is preserved by mapping $\mathcal{M}$ (Lemma \ref{Lemma:CausalityPreserve}), $Lin(\ckt{C}) = Lin(\vc{C})$. (Recall the identity mapping in labels of nodes, used in Theorem \ref{Theorem:isomorphic}).
$ \ckt{T} \in Lin(\ckt{C}), Lin(\ckt{C}) = Lin(\vc{C}) \Rightarrow \ckt{T} \in Lin(\vc{C})$
Hence any event trace produced by the synthesized circuit belongs to the set of feasible traces of the virtual circuit.
\qed

\mysection{Advanced Topics} \label{Sec:advancedtopics}

Now we show how we model some of the advanced features of AHIR framework, using a Petri net pattern, we call \emph{Passive Branch place} and its implementation.
A passive branch is a Petri net place with more than one successors, such that the successors have other non-passive branch predecessor(s) that partly enable the transitions to fire. The passive branch, resolves the conflict by arbitrarily choosing the eligible successor to fire. A \emph{Mutex Place} is just a special case of Passive Branch with initial marking of 1.

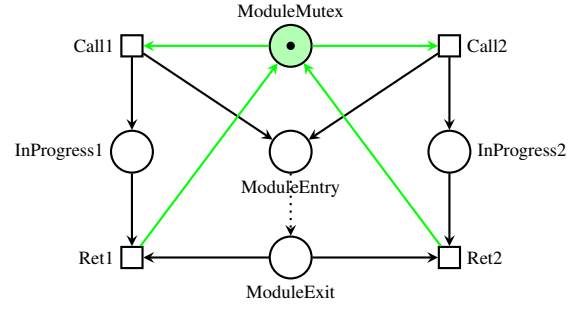
\begin{figure}[]
\centering
\begin{tikzpicture}[scale=.7,transform shape]
\node (mutex) [greenplace,label=ModuleMutex,tokens=1] at (0,2) {};
\node (entry) [tplace,label=below:ModuleEntry] at (0,0) {};
\node (exit) [tplace,label=below:ModuleExit] at (0,-2) {};
\node (caller1) [ttransition,label=left:Call1] at (-3,2) {};
\node (caller2) [ttransition,label=right:Call2] at (3,2) {};
\node (progress1) [tplace,label=left:InProgress1] at (-3,0) {};
\node (progress2) [tplace,label=right:InProgress2] at (3,0) {};
\node (ret1) [ttransition,label=left:Ret1] at (-3,-2) {};
\node (ret2) [ttransition,label=right:Ret2] at (3,-2) {};
\path [greenarc] (mutex.west) to (caller1.east);
\path [greenarc] (mutex.east) to (caller2.west);
\path [arc] (caller1.south) to (progress1.north);
\path [arc] (caller2.south) to (progress2.north);
\path [arc] (progress1.south) to (ret1.north);
\path [arc] (progress2.south) to (ret2.north);
\path [arc] (exit.west) to (ret1.east);
\path [arc] (exit.east) to (ret2.west);
\path [arc] (caller1) to (entry);
\path [arc] (caller2) to (entry);
\path [arc,dotted] (entry) to (exit);
\path [greenarc] (ret1) to (mutex);
\path [greenarc] (ret2) to (mutex);
\end{tikzpicture}

\caption{Petri net model of an arbiter}
\label{Fig:pncallarbiter}
\end{figure}

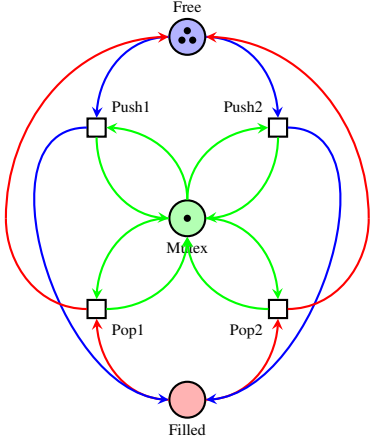
\begin{figure}[]
\centering
\begin{tikzpicture}[scale=.6,transform shape]
\node (mutex) [greenplace,label=below:Mutex,tokens=1] at (0,0) {};
\node (free) [blueplace,label=above:Free,tokens=3] at (0,4) {};
\node (filled) [redplace,label=below:Filled] at (0,-4) {};
\node (push1) [ttransition,label=above right:Push1] at (-2,2) {};
\node (push2) [ttransition,label=above left:Push2] at (2,2) {};
\node (pop1) [ttransition,label=below right:Pop1] at (-2,-2) {};
\node (pop2) [ttransition,label=below left:Pop2] at (2,-2) {};
\path [greenarc,out=90,in=0] (mutex.north) to (push1.east);
\path [greenarc,out=90,in=180] (mutex.north) to (push2.west);
\path [greenarc,out=-90,in=180] (push1.south) to (mutex.west);
\path [greenarc,out=-90,in=0] (push2.south) to (mutex.east);
\path [greenarc,out=180,in=90] (mutex.west) to (pop1.north);
\path [greenarc,out=0,in=90] (mutex.east) to (pop2.north);
\path [greenarc,out=0,in=-90] (pop1.east) to (mutex.south);
\path [greenarc,out=180,in=-90] (pop2.west) to (mutex.south);
\path [bluearc,out=180,in=90] (free.west) to (push1.north);
\path [bluearc,out=0,in=90] (free.east) to (push2.north);
\path [redarc,out=180,in=-90] (filled.west) to (pop1.south);
\path [redarc,out=0,in=-90] (filled.east) to (pop2.south);
\path [bluearc,out=180,in=180] (push1.west) to (filled.west);
\path [bluearc,out=0,in=0] (push2.east) to (filled.east);
\path [redarc,-,out=180,in=-90] (pop1.west) to (-4,0);
\path [redarc,out=90,in=180] (-4,0) to (free.west);
\path [redarc,-,out=0,in=-90] (pop2.east) to (4,0);
\path [redarc,out=90,in=0] (4,0) to (free.east);
\end{tikzpicture}

\caption{Petri net model of communication pipes in AHIR}
\label{Fig:pnpipe}
\end{figure}

\mysubsection{Arbiters for shared module invocation}

Fig. \ref{Fig:pncallarbiter} shows a Petri net model of a call arbiter that manages exclusive access to a module, depicted in green. The Module Exit place is a passive branch, which passes the token to the return transition of the caller that was served.

\mysubsection{Communication channels or pipes}

AHIR framework supports communication channels or pipes \cite{aalrm}. A pipe is associated with a \emph{depth}, which decides the number of buffers available in a pipe.
Fig. \ref{Fig:pnpipe} shows how we model the pipe controller. The part highlighted in green shows exclusiveness of pipe access using a Mutex place. The place named Free, is a passive branch, with an initial marking that represents the pipe depth. The blue arcs show the flow for a push operation that moves a token from the Free place to a passive branch place named Filled. The red arcs show the opposite flow for the pop operations.

\mysubsection{Pipelined loops} \label{Sec:ploops}

AHIR framework has support for generating a pipelined design from an imperative do-while loop, in which each data path operation represents a pipeline stage. The implementation of this involves the following aspects:

\mysubsubsection*{Pipelined interconnect between data path elements}

\begin{figure}[]
\centering
\subfloat[]{
\begin{tikzpicture}[scale=.7,transform shape]
\node (ureq0) [ttransition,label=above:ureq0] at (-2,2) {};
\node (uack0) [ttransition,label=above:uack0] at (0,2) {};
\node (sreq1) [ttransition,label=left:sreq1] at (-1.0,1) {};
\node (sreq2) [ttransition,label=right:sreq2] at (1.0,1) {};
\node (sack1) [ttransition,label=left:sack1] at (-1.0,-1) {};
\node (sack2) [ttransition,label=right:sack2] at (1.0,-1) {};
\node (sackj) [ttransition] at (0,-2) {};
\node (markp) [tplace,tokens=1] at (-3,0) {};
\path [dottedarc,->] (ureq0) to (uack0);
\path [dottedarc,->] (sreq1) to (sack1);
\path [dottedarc,->] (sreq2) to (sack2);
\path [arc] (uack0) to (sreq1);
\path [arc] (uack0) to (sreq2);
\path [arc] (sack1) to (sackj);
\path [arc] (sack2) to (sackj);
\path [arc] (sackj) -| (markp);
\path [arc] (markp) |- (ureq0);
\end{tikzpicture}
}
\subfloat[]{
\begin{tikzpicture}[scale=.7,transform shape]
\node (sreq2) [ttransition,label=left:sreq2] at (0,0.5) {};
\node (sack2) [ttransition,label=above left:sack2] at (0,-1) {};
\node (markp0) [tplace,tokens=1] at (-2.5,1) {};
\node (markp1) [tplace,tokens=1] at (2.5,1) {};
\node (uack0) [ttransition,label=above:uack0] at (-1,3) {};
\node (uack1) [ttransition,label=above:uack1] at (1,3) {};
\node (ureq0) [ttransition,label=above:ureq0] at (-2.5,3) {};
\node (ureq1) [ttransition,label=above:ureq1] at (2.5,3) {};
\node (uackj) [ttransition] at (0,2) {};
\path [dottedarc,->] (sreq2) to (sack2);
\path [dottedarc,->] (ureq0) to (uack0);
\path [dottedarc,->] (ureq1) to (uack1);
\path [arc] (uack0) to (uackj);
\path [arc] (uack1) to (uackj);
\path [arc] (uackj) to (sreq2);
\path [arc] (sack2) -| (markp0);
\path [arc] (sack2) -| (markp1);
\path [arc] (markp0) to (ureq0);
\path [arc] (markp1) to (ureq1);
\end{tikzpicture}
}

\caption{Petri net model of pipelined interconnects: (a) one producer two consumers (b) two producers one consumer}
\label{Fig:loopbody}
\end{figure}
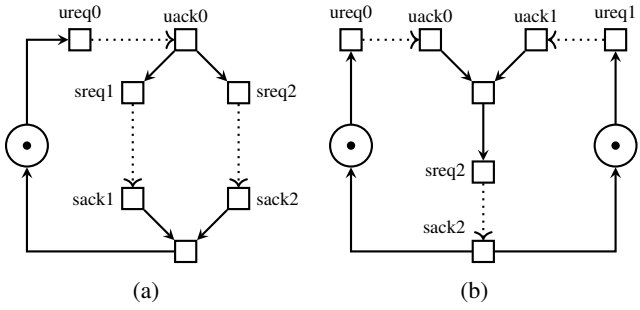

The pipelined interconnect in AHIR is covered in \cite{ee789}. It is based on the interconnect protocol we described in \ref{Def:protocol}. Fig. \ref{Fig:loopbody} shows how AHIR virtual circuit models the interconnect for two representative scenarios. The dotted arcs show implicit dependencies that the data path operator implements. A marked place ensures that the first update (read) request on system initialization is implicitly enabled.

\mysubsubsection*{Pipelined loop entry}

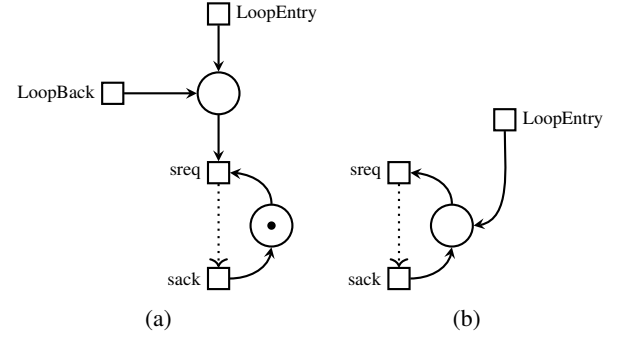
\begin{figure}[]
\centering
\subfloat[]{\label{Fig:loopentry:orig}
\begin{tikzpicture}[scale=.7,transform shape]
\node (entrymerge) [tplace] at (0,-.5) {};
\node (loopback) [ttransition,label=left:LoopBack] at (-2,-.5) {};
\node (loopentry) [ttransition,label=right:LoopEntry] at (0,1) {};
\node (sreq) [ttransition,label=left:sreq] at (0,-2) {};
\node (sack) [ttransition,label=left:sack] at (0,-4) {};
\node (mplace) [tplace,tokens=1] at (1,-3) {};
\path [arc] (loopentry) to (entrymerge);
\path [arc] (loopback) to (entrymerge);
\path [arc] (entrymerge) to (sreq);
\path [arc,out=0,in=-90] (sack) to (mplace);
\path [arc,out=90,in=0] (mplace) to (sreq);
\path [dottedarc,->] (sreq) to (sack);
\end{tikzpicture}
}
\subfloat[]{\label{Fig:loopentry:opt}
\begin{tikzpicture}[scale=.7,transform shape]
\node (loopentry) [ttransition,label=right:LoopEntry] at (2,1) {};
\node (sreq) [ttransition,label=left:sreq] at (0,0) {};
\node (sack) [ttransition,label=left:sack] at (0,-2) {};
\node (mplace) [tplace] at (1,-1) {};
\path [arc,out=0,in=-90] (sack) to (mplace);
\path [arc,out=90,in=0] (mplace) to (sreq);
\path [arc,out=-90,in=0] (loopentry) to (mplace);
\path [dottedarc,->] (sreq) to (sack);
\end{tikzpicture}
}

\caption{(a) Petri net model of loop entry in AHIR virtual circuit and (b) its peephole optimization for non terminating loops}
\label{Fig:loopentry}
\end{figure}

Fig. \ref{Fig:loopentry:orig} shows the Petri net model of entering a pipelined region in AHIR virtual circuits. The LoopEntry transition represents the first time entry into the loop, while LoopBack represents an iterative entry. \sreq represents the sample request of the data path operator that depends on the loop entry.

\mysubsubsection*{Pipelined loop exit controller}

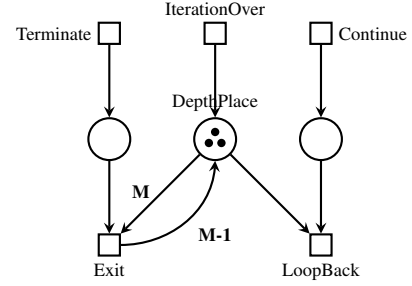
\begin{figure}[]
\centering
\begin{tikzpicture}[scale=.7,transform shape]
\node (lterm) [ttransition,label=left:Terminate] at (-2,2) {};
\node (lbexit) [ttransition,label=above:IterationOver] at (0,2) {};
\node (lcont) [ttransition,label=right:Continue] at (2,2) {};
\node (termplace) [tplace] at (-2,0) {};
\node (depthplace) [tplace,tokens=3,label=above:DepthPlace] at (0,0) {};
\node (contplace) [tplace] at (2,0) {};
\node (lexit) [ttransition,label=below:Exit] at (-2,-2) {};
\node (lback) [ttransition,label=below:LoopBack] at (2,-2) {};
\path [arc] (lterm) to (termplace);
\path [arc] (lbexit) to (depthplace);
\path [arc] (lcont) to (contplace);
\path [arc] (depthplace) to (lexit);
\path [arc] (termplace) to (lexit);
\path [arc] (depthplace) to (lback);
\path [arc] (contplace) to (lback);
\path [arc,out=0,in=-90] (lexit.east) to (depthplace.south);
\node (arcwt) at (-1.4,-1) {\textbf{M}};
\node (arcwt) at (0,-1.8) {\textbf{M-1}};
\end{tikzpicture}

\caption{Petri net model of pipelined loop termination in AHIR}
\label{Fig:pnloop}
\end{figure}

In a pipelined loop, a new iteration of a loop can begin as soon as the loop continuation condition resolves to true, while honoring the data path dependencies. Multiple iterations of a loop may be concurrently active, with different data path elements potentially operating in the context of different iterations. When the loop continuation condition becomes false, the exit controller has to wait till all active iterations are over before exiting.
This involves keeping a count of the iterations in-flight. AHIR specification language lets a designer specify \emph{depth} $M$ of a pipelined loop, which decides the number of bits to be used to implement a counter.
We model this using Petri net, as shown in Fig. \ref{Fig:pnloop} using the Depth place as a passive branch and weighted arcs.

\mysubsubsection*{A peephole optimization for non-terminating pipelined loops}

It is common for hardware pipelines to be perpetually running and in turn many pipelined loops may have the continuation condition to be always true. These loops do not need a loop exit controller described above. We transform the loop entry model of Fig. \ref{Fig:loopentry:orig} into that shown in Fig. \ref{Fig:loopentry:opt} and eliminate the loop exit controller.

\mysubsection{Passive Branch implementation}

\begin{figure}[]
\centering
\begin{tikzpicture}[scale=.5,transform shape]
\node (reso) [combicloud] at (0,0) {Reso};
\node (cnt) [box2] at (3,0) {Cnt\\Reg};
\node (reqs) [box2] at (-10,0) {T2L};
\node (inps) [box2] at (-3,0) {Input\\Reg};
\node (pg) [box] at (-5,-3) {Pulse\\Gen};
\node (d1) [box] at (0,-3) {Delay1};
\node (d2) [box] at (-9,-3) {Delay2};
\coordinate (reqin) at ($(reqs.west)+(-1,.5)$);
\coordinate (tokin) at ($(reqs.west)+(-1,-.5)$);
\coordinate (acks) at ($(reqs.north)+(0,1)$);
\node [anchor=east] (reqinl) at (reqin) {REQS};
\node [anchor=east] (tokinl) at (tokin) {TOKINS};
\node [anchor=south] (acksl) at (acks) {ACKS};
\coordinate (ackvin) at ($(reqs.south)+(.6,0)$);
\coordinate (pulsein) at ($(reqs.south)+(-.6,0)$);
\path [arc,buswidth=m] (reqin) to (reqin-|reqs.west);
\path [arc,buswidth=n] (tokin) to (tokin-|reqs.west);
\path [arc,buswidth=m] (reqs) to (acks);
\path [arc,buswidth=m+n] (reqs.east) -| (pg.north);
\path [buswidth=m+n] ($(reqs.south)+(0,-.75)$) to ($(inps.south)+(0,-.75)$);
\path [arc] (reqs.east -| pg.north) to (inps.west);
\path [arc] (inps.east) to (reso.west);
\path [arc] (reso.east) to (cnt.west);
\path [arc] (pg.east) -| (inps.south);
\path [arc] (pg.east-|inps.south) to (d1.west);
\path [arc] (d1.east) -| (cnt.south);
\path [arc] (reso.south) -- +(0,-.5) -| (ackvin);
\path [arc] (cnt.north) -- +(0,.7) -| (reso.north);
\path [arc] (d1.east -| cnt.south) -- +(0,-1) -| (pulsein);
\path [arc] (pulsein|-d2.west) to (d2.west);
\path [arc] (d2.east) to (pg.west);
\node [anchor=west] (cv) at ($(reqs.east)+(.1,.2)$) {change-vector};
\node [anchor=north] (reqpulse) at (pg.east-|inps.south) {req-pulse};
\node [anchor=west] (ackpulse) at (d1.east-|cnt.south) {ack-pulse};
\node [anchor=west] (ackvector) at ($(ackvin)+(0,-.5)$) {ack-vector};
\node [anchor=west] (resetpulse) at ($(d2.east)+(0,-.2)$) {reset-pg};
\end{tikzpicture}

\caption{Design of the passive branch element}
\label{Fig:pbdesign}
\end{figure}
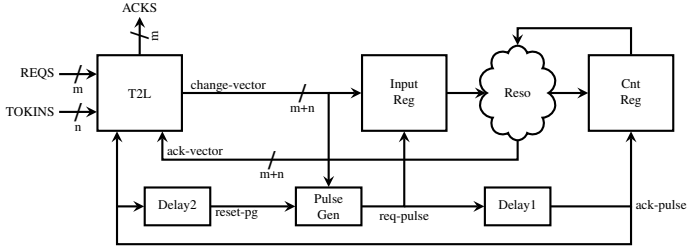

To implement passive branches, we introduce an additional arc by which a transition requests a token from a passive branch.  Fig. \ref{Fig:pbdesign} shows a possible design of a passive branch place. The design supports $m$ request arcs (REQS) and $n$ incoming token arcs (TOKS) as inputs. There are $m$ output arcs (ACKS) that correspond to the REQS. The T2L block is a transition-to-level converter that identifies the change-vector of the inputs. The pulse generator is triggered when there is a request to act upon and becomes insensitive to the input until it receives reset-pg pulse. Input Reg stores a stable snapshot of the input vector and Cnt Reg maintains a counter needed to honor the place capacity and arc weights. Delay1 accounts for the conflict resolution time taken by the combinational logic Reso and the meta stability filtering, while Delay2 holds back resetting of pulse generator until the change vector stabilizes.

We specialize the above design for some specific situations such as, when place capacity and arc weights are 1. In situations where the requests are guaranteed to be mutually exclusive, fundamental mode assumption is used to simplify the circuit.

\mysection{Experimental Results} \label{Sec:results}

We now describe the experimental results of applying our methodology to the synthesis of an asynchronous AES algorithm \cite{rijndael,nist-aes-upd1} implementation and compare them with some of the published results.
Our encryption interface takes a 128-bit \emph{key} and a 128-bit \emph{payload} as an input and it produces a 128-bit \emph{cipher} as an output.

We use pyaes \cite{pyaes} - a Python implementation of AES algorithm to generate the reference results for verification. We use a VHDL test bench for behavioral simulation of the generated RTL to verify the results. We use SPICE simulation to verify the implementation. We use and a test bench written in C++ \footnote{We have created a general purpose component \cite{spicetools} to help write NGSPICE test benches as reactive system that interacts with a circuit under test} and verify the SPICE simulation results with the reference results.

\mysubsection{Design space exploration}

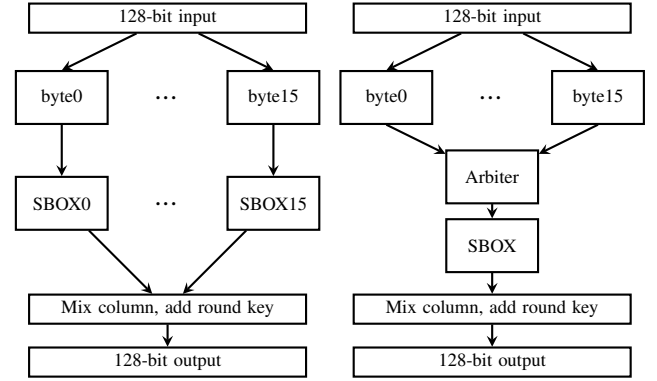
\begin{figure}[]
\begin{subfigure}[c]{.23\textwidth}
\centerline{
\begin{tikzpicture}[scale=.7,transform shape]
\node [box3] (inp) at (0,3.5) {128-bit input};
\node [box] (byte0) at (-2,2) {byte0};
\node [boldtext] (bdots) at (0,2) {\ldots};
\node [box] (byte15) at (2,2) {byte15};
\node [box] (sbox0) at (-2,0) {SBOX0};
\node [boldtext] (sdots) at (0,0) {\ldots};
\node [box] (sbox15) at (2,0) {SBOX15};
\node [box3] (mixcol) at (0,-2) {Mix column, add round key};
\node [box3] (op) at (0,-3) {128-bit output};
\path [arc] (inp) to (byte0.north);
\path [arc] (inp) to (byte15.north);
\path [arc] (byte0) to (sbox0);
\path [arc] (byte15) to (sbox15);
\path [arc] (sbox0) to (mixcol);
\path [arc] (sbox15) to (mixcol);
\path [arc] (mixcol) to (op);
\end{tikzpicture}
}
\caption{Throughput focus: parallel SBOXes}
\end{subfigure}
\begin{subfigure}[c]{.23\textwidth}
\centerline{
\begin{tikzpicture}[scale=.7,transform shape]
\node [box3] (inp) at (0,3.5) {128-bit input};
\node [box] (byte0) at (-2,2) {byte0};
\node [boldtext] (bdots) at (0,2) {\ldots};
\node [box] (byte15) at (2,2) {byte15};
\node [box] (arbiter) at (0,0.5) {Arbiter};
\node [box] (sbox) at (0,-0.80) {SBOX};
\node [box3] (mixcol) at (0,-2) {Mix column, add round key};
\node [box3] (op) at (0,-3) {128-bit output};
\path [arc] (inp) to (byte0.north);
\path [arc] (inp) to (byte15.north);
\path [arc] (byte0.south) to (arbiter);
\path [arc] (byte15.south) to (arbiter);
\path [arc] (arbiter) to (sbox);
\path [arc] (sbox) to (mixcol);
\path [arc] (mixcol) to (op);
\end{tikzpicture}
}
\caption{Area focus: shared SBOX and arbiter}
\end{subfigure}
\caption{Our encryption round data flow}
\label{Fig:EncrDataFlow}
\end{figure}

128-bit AES encryption consists of 10 \emph{rounds} of encryption. \cite{aeswiki} shows a visualization of one round of the encryption flow.  We club the \emph{Byte Sub} and \emph{Shift Row} steps into a \emph{Substitution Box} (SBOX), implemented as a 128:1x8-bit multiplexer. Fig. \ref{Fig:EncrDataFlow} shows our encryption data flow which is implemented in Aa design entry language of the AHIR framework. %

Table \ref{Tab:Confs} describes the design space configurations that we explore and names each of them with a unique identifier for ease of reference.  The configurations that fall under two different design objectives: \emph{area-focused} and \emph{throughput-focused}.  The throughput focused configurations use 16 distinct SBOX instances, while the area focused ones use a single SBOX instance, shared using an arbiter.

AHIR framework supports describing pipelined designs in the form of looping constructs in the Aa language. The configuration \texttt{Our.pl} uses each encryption round as a pipeline stage, while \texttt{Our.16x} is a non-pipelined, combinational design. Among the area focused configurations, we explore different cascaded arrangements of arbiters such as a single 16:1 arbiter instance and cascaded usage of 2:1, 4:1 and 8:1 arbiters, identified with respective names.

\mysubsection{Simulation results}

\begin{table}[]
\centering
\caption{Description of the configurations explored.}
\begin{tabular}{@{}lllll@{}l}
\hline
\textbf{Config}
&
\textbf{Objective}
&
\textbf{SBOXes}
&
\textbf{Arbiter}
&
\textbf{Pipelined}
\\\hline
Our.16x
&
Throughput
&
16
&
-
&
-
\\\hline
Our.pl
&
Throughput
&
16
&
-
&
Each round as a stage
\\\hline
Our.161
&
Area
&
1
&
16:1
&
-
\\\hline
Our.21
&
Area
&
1
&
2:1
&
-
\\\hline
Our.41
&
Area
&
1
&
4:1
&
-
\\\hline
Our.81
&
Area
&
1
&
8:1
&
-
\\\hline
\end{tabular}
\label{Tab:Confs}
\end{table}

\begin{table*}[]
\centering
\caption{Various metrics gathered from surveyed AES implementations.}
\setlength{\tabcolsep}{4pt}
\begin{tabular}{@{}rrrrrrrrrrrrrrrrrrrrrrrr@{}}
\textbf{Ref} & \textbf{Year} & \textbf{Tech} & \textbf{Vdd} & \textbf{Thrpt} & \textbf{Power} & \textbf{Energy} & \textbf{Thrpt/} & \textbf{Thrpt/} & \textbf{Thrpt/} & \textbf{EDP} & \textbf{ED\textsuperscript{2}P} & \textbf{EDA} & \textbf{EDA\textsuperscript{2}} & \textbf{Area} & \textbf{GEs}
\\
\textbf{} & \textbf{} & \textbf{} & \textbf{} & \textbf{} & \textbf{} & \textbf{per op} & \textbf{Energy} & \textbf{Power} & \textbf{Area} & \textbf{} & \textbf{} & \textbf{} & \textbf{} & \textbf{} & \textbf{}
\\
 &  & nm & V & Mbps & \SI{}{\micro\watt} & nJ & Mbps/ & \SI{}{Mbps/} & Mbps/ & \SI{}{\nano\joule\second} & \SI{}{\nano\joule\second\squared} & \SI{}{\nano\joule.\micro\second.} & \SI{}{\nano\joule.\micro\second.} & \SI{}{\milli\meter\squared} & 
\\
 &  &  &  &  &  &  & nJ & \SI{}{\micro\watt} & \SI{}{\milli\meter\squared} &  &  & \SI{}{\milli\meter\squared} & \SI{}{\milli\meter^4} &  & 
\\\hline
\cite{Feldhofer-04} & 2004 & 350 & 1.5 & 1.3e-2 & 1.2e+1 & 1.2e+2 & 1.0e-4 & 1.1e-3 & - & 1.2e+0 & 1.2e-2 & - & - & - & 3.6e+3
\\\hline
\cite{Feldhofer-05} & 2005 & 350 & 1.5 & 9.9e+0 & 3.6e+3 & 4.6e+1 & 2.1e-1 & 2.8e-3 & 4.0e+1 & 6.0e-4 & 7.8e-9 & 1.5e+2 & 3.8e+1 & 2.5e-1 & 3.4e+3
\\\hline
\cite{Hamalainen} & 2006 & 130 & 1.2 & 1.2e+2 & 5.6e+3 & 5.9e+0 & 2.1e+1 & 2.2e-2 & - & 6.2e-6 & 6.6e-12 & - & - & - & 3.1e+3
\\\hline
\cite{Good} & 2010 & 130 & 0.75 & 3.6e-2 & 6.9e-1 & 2.5e+0 & 1.5e-2 & 5.2e-2 & 1.7e+0 & 8.8e-3 & 3.1e-5 & 1.8e+2 & 3.9e+0 & 2.1e-2 & 5.5e+3
\\
\hline
\end{tabular}
\label{Tab:Comparison}
\end{table*}

\begin{figure*}[]
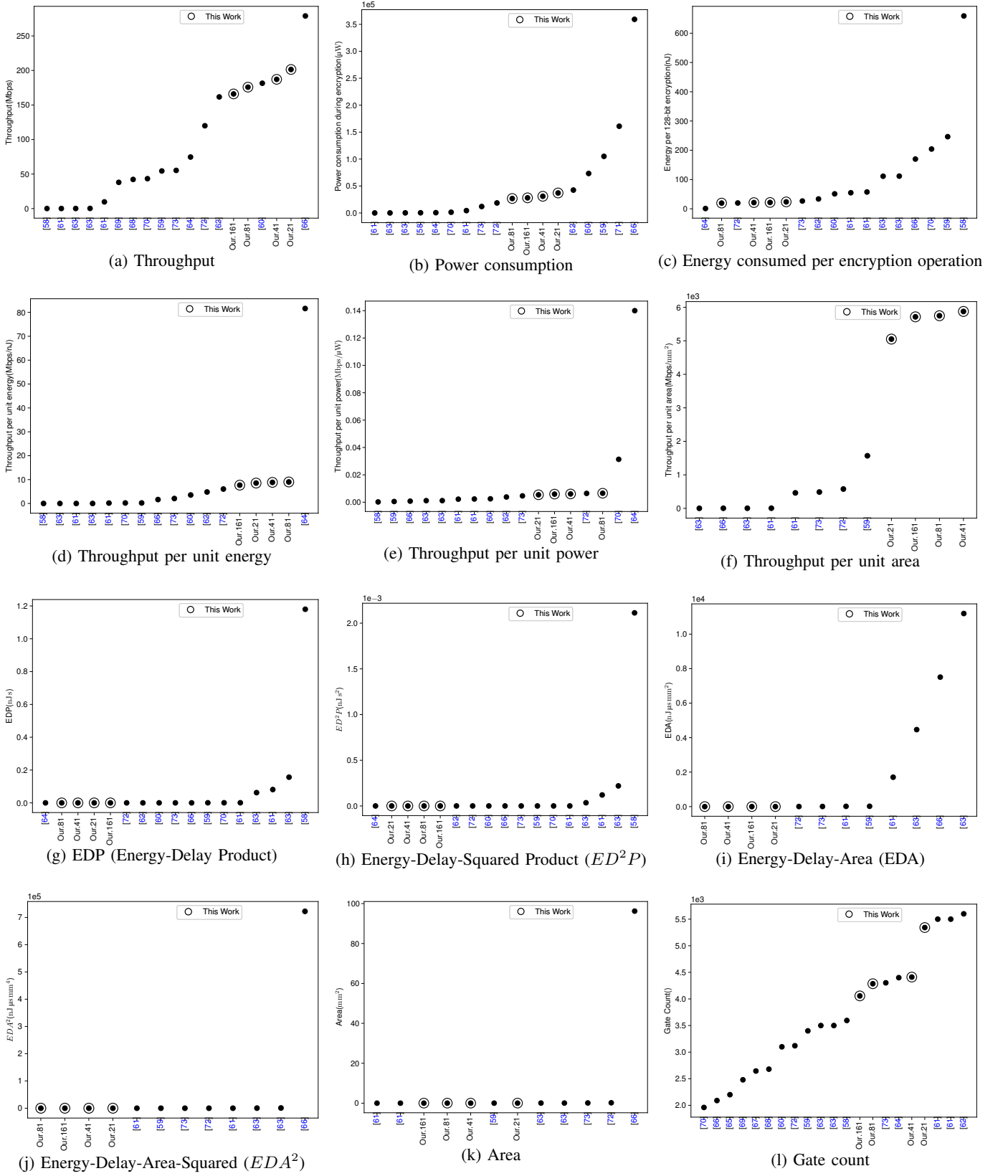


\begin{subfigure}[c]{.33\textwidth}
\resizebox{1\textwidth}{!}{\input{N\_Throughput.pgf}}
\vspace{-2em}
\caption{Throughput}
\end{subfigure}
\vspace{.3em}
\begin{subfigure}[c]{.33\textwidth}
\resizebox{\textwidth}{!}{\input{N\_powerAtMaxF.pgf}}
\vspace{-2em}
\caption{Power consumption}
\end{subfigure}
\vspace{.3em}
\begin{subfigure}[c]{.33\textwidth}
\resizebox{1\textwidth}{!}{\input{N\_energy.pgf}}
\vspace{-2em}
\caption{Energy consumed per encryption operation}
\end{subfigure}
\vspace{.3em}
\begin{subfigure}[c]{.33\textwidth}
\resizebox{1\textwidth}{!}{\input{N\_MbpsPernJ.pgf}}
\vspace{-2em}
\caption{Throughput per unit energy}
\end{subfigure}
\vspace{.3em}
\begin{subfigure}[c]{.33\textwidth}
\resizebox{\textwidth}{!}{\input{N\_MbpsPeruW.pgf}}
\vspace{-2em}
\caption{Throughput per unit power}
\end{subfigure}
\vspace{.3em}
\begin{subfigure}[c]{.33\textwidth}
\resizebox{1\textwidth}{!}{\input{N\_MbpsPerSqmm.pgf}}
\vspace{-2em}
\caption{Throughput per unit area}
\end{subfigure}
\vspace{.3em}
\begin{subfigure}[c]{.33\textwidth}
\resizebox{1\textwidth}{!}{\input{N\_EDP.pgf}}
\vspace{-2em}
\caption{EDP (Energy-Delay Product)}
\end{subfigure}
\vspace{.3em}
\begin{subfigure}[c]{.33\textwidth}
\resizebox{1\textwidth}{!}{\input{N\_ED2P.pgf}}
\vspace{-2em}
\caption{Energy-Delay-Squared Product ($ED^2P$)}
\end{subfigure}
\vspace{.3em}
\begin{subfigure}[c]{.33\textwidth}
\resizebox{\textwidth}{!}{\input{N\_EDA.pgf}}
\vspace{-2em}
\caption{Energy-Delay-Area (EDA)}
\end{subfigure}
\vspace{.3em}
\begin{subfigure}[c]{.33\textwidth}
\resizebox{\textwidth}{!}{\input{N\_EDA2.pgf}}
\vspace{-2em}
\caption{Energy-Delay-Area-Squared ($EDA^2$)}
\end{subfigure}
\vspace{.3em}
\begin{subfigure}[c]{.33\textwidth}
\resizebox{1\textwidth}{!}{\input{N\_Area.pgf}}
\vspace{-2em}
\caption{Area}
\end{subfigure}
\vspace{.3em}
\begin{subfigure}[c]{.33\textwidth}
\resizebox{1\textwidth}{!}{\input{N\_GEs.pgf}}
\vspace{-2em}
\caption{Gate count}
\end{subfigure}
\vspace{.3em}

\caption{Plots of normalized figures of merit for various AES implementations and our study, excluding our throughput focused configurations}
\label{Fig:NComparisonPlots}

\end{figure*}

\begin{table}[]
\centering
\caption{Dynamic and idle power consumption}
\begin{tabular}{@{}rrrr@{}}
\hline
\textbf{Config} &
\textbf{Dynamic} &
\textbf{Idle} &
\textbf{Dy-to-idle}\\
&\textbf{Power}&\textbf{Power}&\textbf{Ratio}\\
&(\SI{}{\micro\watt})&(\SI{}{\micro\watt})&
\\\hline
Our.16x&
2.6e+4 &
1.4e+1 &
1.9e+3
\\\hline
Our.pl&
2.4e+5 &
1.6e+2 &
1.5e+3
\\\hline
Our.161&
7.3e+3 &
3.2e+0 &
2.3e+3
\\\hline
Our.21&
9.6e+3 &
4.5e+0 &
2.1e+3
\\\hline
Our.41&
8.0e+3 &
3.8e+0 &
2.1e+3
\\\hline
Our.81&
6.9e+3 &
3.7e+0 &
1.9e+3
\\\hline

\end{tabular}
\label{Tab:IdlePower}
\end{table}

We restrict the SPICE simulation to 1 round of encryption for non-pipelined configurations and 2 rounds for the pipelined configuration. We extrapolate the results to 10 rounds of encryption. Computation of metrics we present, is extrapolated to 10 rounds, from SPICE the measurements.
For the pipelined configuration, the SPICE measurements are based on a two-stage pipeline (two encryption rounds) running over two inputs. The gate count and area is extrapolated to ten-stage pipeline. The throughput is computed, basis the time duration between two outputs. For energy and power computations, the results are extrapolated assuming the pipeline is not starved for inputs.

Comparing results across the diversity of technology nodes is challenging. Stillmaker et al. articulate this well \cite{stillmaker}. We use their DeepScale tool \cite{deepscale}, to normalize the time, power and area to 130 nm node used in our experimentation. Fig. \ref{Fig:NComparisonPlots} shows sorted scatter plots of normalized FoMs. We exclude the throughput focused configurations from the plots since their performance and resource metrics are in a different range than the area focused configurations. In Table \ref{Tab:IdlePower} we list the idle power consumption and the ratio of dynamic to idle power consumption for our experiments.

Some observations:
\begin{inparaenum}
\item The results establish the feasibility of our asynchronous circuit synthesis methodology and show that the results are competitive.
\item The study also demonstrates the agility of our framework to explore the design space by creating and benchmarking various configurations with area-throughput trade offs.
\item Our throughput-focused configurations produce best in class throughput, although with higher area cost and power consumption.
\item Our area-focused configurations use a gate count and area that is comparable with other results, but produce better throughput than the most.
\item Both area and throughput focused configurations perform well in FoMs throughput-per-unit-area, energy-delay-area and energy-delay-area-square.
\item We observe that the idle power consumption of the asynchronous circuits produced is approximately 2000 times lower than the dynamic power consumption, which highlights a significant benefit of the asynchronous circuits.
\end{inparaenum}

\mysection{Future Work} \label{Sec:futurework}

\begin{inparaenum}
\item To enable our methodology to be a part of a larger VLSI ecosystem, such as chiplet based systems, which may use a mix of asynchronous and synchronous components, it will be useful to build protocol adapters to interoperate with synchronous components. Interfacing with external memory (RAM) is often going to be a requirement in many chip designs.
\item Post-implementation tuning of time delays, to meet the timing constraints, is a manual step involved in our flow. Our delay elements are designed with a \emph{fixed} and a \emph{variable} delay component, such that the variable component can be changed post-implementation, using the constant bits embedded in the design. However, currently we do not have an estimation model for the delay that can take benefit of this ability. The delay required is a function of the operator(s) that constitute the combinational logic of a pipeline stage, the data width, the placement and routing and the technology node. It is possible to empirically arrive at a prediction model that predicts the delay requirements as a range of upper and lower bounds. The lower bound can be used to specify the fixed component of the delay and the slack between the lower and upper bound to specify the variable component. This can speed up the process of delay tuning and potentially automate it completely.
\item If we can make placement and routing aware of the control and data path distinction and can give higher priority to data path nets, it will reduce the delay requirements, leading to area and energy saving as well as potentially reduction in the latency.
\item Off-the-shelf cell libraries often do contain delay elements that make use of buffers with larger input areas to produce delays. However, a bundled data design we often need to cascade multiple such elements, to meet the delay constraints. Besides increased area, these also increase the energy consumption by the delay elements. The physical design aspect of these elements can be explored to create delay elements with suitable characteristics that meet the delay requirements with fewer stages, thus reducing the area and energy costs.
\item Our experimental results are based on typical operating conditions of supply voltage and temperature. Further exploration may be taken up with greater variation in these parameters to assess their impact on various figures of merit. We may also explore more classes of applications to assess the suitability of the methodology and asynchronous paradigm.
\end{inparaenum}

\mysection{Conclusion} \label{Sec:conclusion}

\begin{inparaenum}
\item We demonstrated a high-level synthesis framework for asynchronous circuits with a flow from imperative programs to RTL that can feed to any off-the-shelf ASIC flow built for the synchronous design.
Our framework contains a static timing analysis tool to tune the delays to meet the timing constraints, which we showed to be feasible, necessary and sufficient. We proposed some ways to minimize or potentially eliminate the manual intervention for the timing closure.
\item We formalized the problem of synthesis of asynchronous circuits using a restricted subset of Petri nets that AHIR framework builds. Compared to the established methods like STGs, our Petri net formulation is simpler because we can isomorphically map it to a synthesized circuit, without involving complete state coding problem like STGs.
\item Using our methodology, we designed a circuit for AES encryption. We explored the design space, with different area-latency-energy trade offs. We performed transistor level simulation using SPICE and compared the metrics gathered with various published results. We find our results to be competitive; particularly the energy efficiency and idle power consumption show that such methodology could fit the use cases where these metrics matter.
\end{inparaenum}

\balance
\bibliographystyle{IEEEtran}
\bibliography{citations}

\end{document}


\newcommand\swtab[3]{
\item \emph{#1}: #2
URL: \url{#3}
}
\appendix \label{Sec:appendix}
\appendixstyle

\mysection{List of software components developed}

The following software artifacts were developed as part of this work and are publicly available.

\begin{inparaenum}
\swtab{AHIR Asynchronous framework}
{
The main framework described in this paper, consisting of
\begin{inparaenum}
\item VHDL code generator from the AHIR virtual circuit representation.
\item A library of asynchronous components implemented in VHDL.
\item A static timing analyzer for the asynchronous circuits.
\item Netlist models of some key components and specification for checking some properties using the Spin model checker.
\end{inparaenum}
}
{https://github.com/mayureshw/ahirasync}

\swtab{Promela code generator from a netlist}
{
A utility to specify digital circuits and properties to be verified using the Spin model checker. Consists of
\begin{inparaenum}
\item A netlist format to express a digital circuit and notation to express LTL properties.
\item A converter from netlist to Promela code.
\item A library that encodes gate and wire behaviors in Promela.
\item A Make based flow for verification of LTL properties.
\end{inparaenum}
}
{https://github.com/mayureshw/netlist2promela}

\swtab{C++ interface with ngspice simulator}
{
A library that makes writing SPICE test benches easy and provides debugging help. Consists of
\begin{inparaenum}
\item A digital interface around SPICE which is an analog circuit simulator
\item The interface lets club nets and identify them as vectors
\item Provides a state machine interface to write test benches, that can track states of interest and respond with events on the input ports of the SPICE circuit.
\item Allows playback of a simulation log, while logging changes to the values of nets or vectors of nets of interest.
\item Detects nets in metastable states for a duration longer than a specifiable threshold.
\end{inparaenum}
}
{https://github.com/mayureshw/spicetools}

\swtab{A virtual circuit simulator with randomization of the order of events}
{
A simulator for the virtual circuit representation in AHIR system, built on top of a Petri net simulator and a complex event processing (CEP) tool. Can perform simple behavioral simulation as well as randomized simulation where unordered events are processed in a random order or using a specifiable stochastic model. Randomization helps detect subtle race condition related defects with a quantifiable measure of confidence.
}
{https://github.com/mayureshw/vctools}

\swtab{A Petri net simulator}
{
A general purpose Petri net simulator, with ability to impose a given stochastic or random order on the events that have no causal relation. This is used as a core component by the virtual circuit simulator.
}
{https://github.com/mayureshw/petrisimu}

\swtab{A CEP tool}
{
A verifier for event sequence properties on a stream of events. Consists of a specification mechanism for properties over event intervals and a runtime interface to verify those properties on a stream of events. This used by the virtual circuit simulator.
}
{https://github.com/mayureshw/ceptool}

\swtab{C++ interface to XSB Prolog}
{
A general purpose two way bridge between C++ and XSB Prolog system. Lets create and send goals from C++ to Prolog system and read and bind results from Prolog system to C++ data structures. This is used by the CEP tool.
}
{https://github.com/mayureshw/xsbcppif}

\end{inparaenum}